\documentclass[11pt]{article}
\usepackage[a4paper,lmargin={1.2cm},rmargin={1.2cm},tmargin={2cm},bmargin =
{2cm}]{geometry}

\usepackage{amsfonts}
\usepackage{amsmath}

\usepackage{graphicx}

\usepackage{cite}
\usepackage[colorlinks=true,linkcolor=blue,citecolor=blue,urlcolor=blue]{hyperref}

\usepackage{caption}
\usepackage{xcolor}

\newcommand{\rme}{\mathrm{e}}
\newcommand{\rmi}{\mathrm{i}}
\newcommand{\rmd}{\mathrm{d}}
\newcommand{\rmv}{\mathrm{v}}

\renewcommand{\qquad}{\hspace*{25pt}}

\begin{document}

\sloppy

\begin{center}
{\Large\bf Axionlike dark-matter winds driven by galactic baryon redistribution}\\

\vspace*{2.5mm}
{A.V. Nazarenko\footnote{e-mail: nazarenko@bitp.kyiv.ua}}\\

\vspace*{1.5mm}
{\small Bogolyubov Institute for Theoretical Physics of NAS of Ukraine\\
14b, Metrolohichna Str., Kyiv 03143, Ukraine}
\end{center}

%% Abstract
\abstract{We examine solutions of the hydrodynamic equations
for dark matter (DM) modeled as a Bose--Einstein condensate (BEC)
with axion-like interaction, forming a spherically symmetric halo
in dwarf galaxies. Small perturbations and macroscopic decoherence
of the BEC DM are driven by variations in the local gravitational
background induced by subgalactic baryonic redistribution. Focusing
on the core region of a dwarf galaxy, we analyze three astrophysical
scenarios parametrized by observational data for NGC 2366:
(i)~expansion of a gaseous shell mimicking stellar explosions,
(ii)~collapse of a shell modeling star formation, and (iii)~inward
contraction of a stellar cluster driven by dynamical friction within
a gaseous environment. Our linearized hydrodynamic simulations
reveal that while central DM density variations remain within
a subtle 0.01\% margin, the induced DM winds are extremely
suppressed, reaching velocities of only up to several meters
per second. This striking suppression challenges the observational
prospect of detecting dark-wind signatures.
To address the long-term cumulative impact of multiple baryonic
events, we establish a quantum excitation framework with a discrete
spectrum under natural Robin boundary conditions.
In the inhomogeneous DM halo background, which includes
an unstable phase region, the lowest long-wavelength radial mode
acts as a false vacuum state, experiencing non-viscous quantum
relaxation and decaying over a characteristic lifetime of
$6.8\times10^7$ years. This mechanism triggers a universal
exponential damping of the system’s evolutionary operator, while
nonlinear three-mode parametric energy exchange, rather than
standard open-continuum Beliaev damping, emerges in the next
order of perturbation theory to govern the redistribution of
stable quasiparticle excitations.}

%% %% %% %% %% %% %% %% %% %% %% %% %% %% %% %% %% %% %% %% %% %% %% %% %% %% %% %%
\section{Introduction}\label{S1}

The paradigm of ultralight axion-like particles (ALPs) with masses of
$m\sim 10^{-22}~\text{eV}\,c^{-2}$ has emerged as a compelling alternative to
standard cold dark matter (CDM), offering a natural quantum-mechanical resolution
to small-scale cosmological puzzles such as the core-cusp and missing satellite
problems~\cite{Bal83,BGZ84,UTB02,FMT08,Bor16,KM17}. While the theoretical origins
of ALPs are deeply rooted in string-theoretic compactifications, their cosmological
production mechanism is conceptually projected from quantum chromodynamics (QCD).
Specifically, these coherent axion states are assumed to be nonthermally generated
via the vacuum misalignment mechanism~\cite{PWW,AS83,DF83} and the decay of topological
defects in the early Universe~\cite{Dav86}.

Although early cosmological models -- commonly called fuzzy dark matter (FDM)
-- treated the dark matter (DM) halo as a non-interacting, free scalar field
governed solely by linear quantum pressure~\cite{Sin,Mar16,OW16,Ter19}, realistic
galactic configurations require the inclusion of
self-interactions~\cite{Wit80,KT94,Ch2018,Wil18,SH2018,BZ19,Nambo24} or mixed DM
scenarios~\cite{Marsh2014}. To bridge the scale gap between particle physics
and astrophysics, we project the nonperturbative features of large-$N$ QCD
chiral dynamics~\cite{Wit80,VV80,Cor16} onto the ultralight ALP sector.
Treating the axion mass $m$ and the decay constant $f_{a}$ as independent
macroscopic parameters of an effective field theory allows the nonlinear
interaction to manifest as a periodic, nonperturbative cosine-like potential.
The resulting macroscopically coherent Bose--Einstein condensate (BEC) exhibits
complex wave properties that depart from the free-field baseline,
thereby rendering galactic halos as self-gravitating quantum fluids
whose static distributions have been extensively
studied~\cite{Lee96,SY09,SY12,Dav13,Dav15,GHP15,Harko2011,CH20,Kun2020}.

A pivotal consequence of nonlinear ALP self-interactions is the emergence of
a multi-phase halo structure~\cite{Ch2017,Ch25x,GKN20}. As established
in~\cite{Naz2025}, evaluating the grand canonical potential of a self-gravitating
BEC with the cosine-like potential reveals alternating zones of mechanical
stability and instability on the parametric $p-\mu$ phase diagram. In low-density
regions corresponding to the spatial tails, the medium possesses an anomalous
compressibility in which the squared speed of sound becomes strictly negative
($c_s^2<0$). Thermodynamically, this domain marks the onset of spinodal
decomposition. Far from signaling a breakdown
of the hydrodynamic description, this unstable phase triggers local BEC
fragmentation into high-density clumps (such as axion miniclusters~\cite{KT93})
or foam-like structures. Crucially, while a massive solitonic core, exceeding
$\sim 12\%$ of the total mass, can gravitationally shield the halo against
minor perturbations~\cite{Naz2025}, real subgalactic structures remain
exposed to ongoing external disruptions.

In DM-dominated systems, such as dwarf galaxies, substantial baryonic
redistribution driven by supernova feedback, stellar winds, or starburst
activities acts as a time-dependent external gravitational perturber.
While baryon-DM interactions have been widely studied within collisionless
cold~\cite{MSch15,Sch23} and warm DM models~\cite{RG05,Mash06,GM06,Gov10,Gov12},
the dynamic response of a coherent BEC fluid remains comparatively
obscure~\cite{Ch12,H2019,Ch2020,Ferreira,HM2022}. This baryonic perturbation
inevitably breaks the global coherence of the stationary BEC distributions,
exciting unique acoustic collective modes of ``dark sound'' \cite{Bal22} and
driving local phase slips that generate macroscopic hydrodynamic currents,
which we call DM winds. However, evaluating the forced response of a medium
that partially straddles an unstable phase zone ($c_s^2<0$) introduces severe
theoretical challenges, as the long-wavelength sector is inherently prone
to Jeans-like and self-interaction destabilization.

In this work, we extend the static thermodynamic framework established in~\cite{Naz2025}
into the fully dynamical regime, investigating the nonequilibrium relaxation of
the BEC DM halo in the reference dwarf galaxy NGC 2366. Using observational
data~\cite{Oh08}, we model three distinct baryonic scenarios in Section~\ref{S3}:
the rapid expansion of a supernova-driven gas shell, the free-fall collapse of a
star-forming cloud, and the migration of a stellar cluster. To complete the latter
scenario, we develop an extended concentric-shell formulation of Chandrasekhar’s
dynamical friction, generalizing the classical pointlike drag approximation~\cite{Chandra43}
to a symmetric two-shell geometry, which reveals distinct features compared
to other friction models~\cite{Rud71,Reph80,Ostr,Tre84,Kim09,Lee13,Des22,Lan20,Buh23,
Lee25,Gor25,BEK19}.
We demonstrate that while typical standing-wave constraints stabilize short-term
hydrodynamic flows, a comprehensive description of cumulative baryonic events
requires a transition to an autonomous quantum picture under natural Robin
boundaries. Within this approach in Section~\ref{S4}, we resolve the non-trivial
expansion and show that the unstable long-wavelength radial sector behaves as
a metastable false vacuum~\cite{Col77,Sh96,YSL}. Quantum tunneling across the
self-generated pressure barrier dictates a non-viscous relaxation mechanism, leading
to a universal exponential decay of perturbations via nonlinear three-mode parametric
wave mixing rather than standard continuum Beliaev damping~\cite{LP80,Bel2,MVS,CB09,P15}.

%% %% %% %% %% %% %% %% %% %% %% %% %% %% %% %% %% %% %% %% %% %% %% %% %% %% %% %%
\section{The DM model equations}\label{S2}

We consider the Bose--Einstein condensate (BEC) of nonrelativistic particles with
mass~$m$, described by the macroscopic complex wave function $\psi(r,t)$ of
time~$t$ and radial coordinate~$r$ inside a three-dimensional ball $B(r\leq R)$.
The incorporation of the axion-like self-interaction~$V_{\rm int}$ and the potential
of gravitational interaction~$V_{\rm gr}$ leads to the Hamiltonian functional:
\begin{eqnarray}
&&H=4\pi\int_0^R\left[\frac{\hbar^2}{2m}|\partial_r\psi|^2
+mV_{\rm gr}|\psi|^2+V_{\rm int}\right] r^2\,\rmd r,
\label{G1}\\
&&V_{\rm int}=\frac{U}{\rmv}\left[1-\cos{\left(\sqrt{\rmv}|\psi|\right)}\right]
-\frac{U}{2}|\psi|^2,
\label{SI1}\\
&&\Delta_r V_{\rm gr}=4\pi Gm|\psi|^2,
\label{Poi}
\end{eqnarray}
where $|\psi|^2=\psi\overline{\psi}$; $\overline{\psi}$ is the complex conjugate of $\psi$;
$G$ is Newton's constant.

The axion-like interaction (\ref{SI1}) is chosen in analogy with \cite{PQ77,Wit80},
assuming the connection between the axion field $\varphi$ with the nonrelativistic
wave function $\psi$ via the relation $|\varphi|^2=(\hbar^2/m) |\psi|^2$~\cite{Ch2018}.
Other effective axion potentials also appear in \cite{VV80,Guzman,Cor16,Ch2017}.

The self-interaction (\ref{SI1}) is characterized by two constants, $U$
and $\rmv$, which have dimensions of energy and volume. In the particle
physics~\cite{VV80,Cor16}, they are related with the axion mass $m$ and
decay constant $f_{\text{a}}$ as $U=mc^2$, $\rmv=\hbar^3c/(mf^2_{\text{a}})$.
This mapping of relativistic definitions onto non-relativistic parameters
is convenient for scaling and comparison between different models.

The Poisson equation (\ref{Poi}), along with its solution, is based on the radial
part of the Laplace operator $\Delta_r$ and its inverse $\Delta^{-1}_r$
with respect to the variable $r$:
\begin{eqnarray}
&&\hspace*{-5mm}
\Delta_rf(r)=\partial^2_rf(r)+\frac{2}{r}\,\partial_rf(r),
\label{Dlt}\\
&&\hspace*{-5mm}
\Delta^{-1}_rf(r)=-\frac{1}{r}\int_0^rf(s)s^2\rmd s-\int_r^{R}f(s)s\rmd s.
\label{InvDelta}
\end{eqnarray}

The Gross--Pitaevskii equation is derived as
\begin{eqnarray}
&&\rmi\hbar\frac{\partial\psi(r,t)}{\partial t}=\frac{\delta H}{\delta\overline{\psi}(r,t)},
\end{eqnarray}
taking into account the integration measure and the rules of functional differentiation
for functions of the same time~$t$:
\begin{equation}
\frac{\delta\psi(r^\prime,t)}{\delta\psi(r,t)}=\frac{1}{4\pi rr^\prime}\delta(r-r^\prime),
\quad \frac{\delta\psi(r^\prime,t)}{\delta\overline{\psi}(r,t)}=0.
\end{equation}

We find explicitly that
\begin{equation}
\rmi\hbar\partial_t\psi=-\frac{\hbar^2}{2m}\Delta_r\psi+mV_{\rm gr}\psi
+\frac{U}{2}\left[j_0\left(\sqrt{\rmv}|\psi|\right)-1\right]\psi,
\label{GP1}
\end{equation}
where we refer to the spherical Bessel functions:
\begin{equation}
j_l(z)=z^l\left(-\frac{1}{z}\frac{\rmd}{\rmd z}\right)^l\frac{\sin{z}}{z}.
\end{equation}

Thus, the generalized Gross--Pitaevskii equation (\ref{GP1}) together with
the Poisson equation (\ref{Poi}) solved using Eq.~(\ref{InvDelta}) -- GPP
equations --  determine the dynamics of the order parameter $\psi$ in
our model of the BEC DM.

Validating the coupled GPP equations (\ref{GP1}), (\ref{Poi})
against empirical phenomena is a challenging task, even when
baryon feedback is minimized. The obstacles extend beyond nonlinear
mathematical complexities and require a rigorous definition
of the physical system. In our case, the parameters determining
the initial wave function $\psi(r)$ depend on the choice of the
galactic environment.

We intend to minimize the number of parameters after their
nondimensionalization to ensure clarity of solutions to the
equations formulated for the selected DM-dominated galaxy.

Thus, we perform the replacement of model parameters $(m,U,\rmv)\mapsto(r_0,A,f_{\rm a})$:
\begin{eqnarray}
&&r_0=\frac{\hbar}{\sqrt{mU}},\quad
A=\frac{8\pi G\hbar^2m}{U^2\rmv},\quad
f_{\rm a}=\sqrt{\frac{\hbar^3c}{m\rmv}};
\label{par1}\\
&&m=\left[\frac{\hbar^5c}{8\pi G}\frac{A}{f_{\rm a}^2r_0^4}\right]^{1/4},
\quad
U=\frac{\hbar^2}{mr_0^2},\quad
\rmv=\frac{\hbar^3c}{mf_{\rm a}^2}.
\label{par2}
\end{eqnarray}
The relative strength of gravitational interaction $A$ is a dimensionless
parameter, while $r_0$ is the distance scale, and $f_{\rm a}$ in energy
units determines the decay constant of ALPs.

Evaluating these relations at characteristic galactic physical scales
yields the mass estimate:
\begin{equation}
m=0.17745\times10^{-22}~\text{eV}~c^{-2}~\left[\frac{A}{10^{-3}}\right]^{1/4}
\left[\frac{r_0}{1~\text{kpc}}\right]^{-1}
\left[\frac{f_\text{a}}{10^{19}~\text{eV}}\right]^{-1/2}.
\end{equation}

We also introduce the scales of time and energy:
\begin{equation}
t_0=\frac{mr_0^2}{\hbar},\quad
\varepsilon_0=\frac{\hbar^2}{2mr_0^2}.
\end{equation}

Parametrizing the space and time variables by $\xi$ and $\tau$:
\begin{equation}
r=r_0\xi,\quad R=r_0\xi_B, \quad t=t_0\tau,
\end{equation}
we define the dimensionless wave function:
\begin{equation}
\phi(\xi,\tau)=\sqrt{\rmv}\,\psi(r_0\xi,t_0\tau).
\end{equation}

Combining, we arrive at the set of dimensionless equations:
%
%\begin{eqnarray}
\begin{subequations}\label{feqs}
\begin{align}
&2\rmi\partial_{\tau}\phi=-\Delta_\xi\phi+A\Phi_{\rm gr}\phi
+\left[j_0\left(|\phi|\right)-1\right]\phi,
\label{feq1}\\
&\Phi_{\rm gr}(\xi,\tau)=\Phi_0(\tau)
+\int_0^{\xi}\left(\frac{1}{s}-\frac{1}{\xi}\right)|\phi(s,\tau)|^2s^2\rmd s,
\label{feq2}
\end{align}
\end{subequations}
%\end{eqnarray}
%
where the term
\begin{equation}\label{Phi0}
\Phi_0(\tau)\equiv-\int_0^{\xi_B}|\phi(s,\tau)|^2s\,\rmd s
\end{equation}
defines the gravitational potential at the origin.

Although a complex solution $\phi$ to Eqs.~(\ref{feqs})
is governed by the dimensionless constant~$A$, the physical
characteristics also depend on the dimensional
$r_0$ and $f_{\rm a}$. Defining the amplitude~$\chi$:
\begin{equation}\label{Mad}
\phi(\xi,\tau)=\chi(\xi,\tau)\, \rme^{\rmi\theta(\xi,\tau)},
\end{equation}
the local DM mass density is $\rho=\rho_0\chi^2$, where $\rho_0\equiv m/\rmv$,
and the number of DM particles contained under a sphere of radius $\xi=r/r_0$ is
determined, omitting the factor $4\pi r_0^3/\rmv$, by the integral:
\begin{equation}\label{Nt}
n(\xi,\tau)=\int_0^\xi\chi^2(s,\tau)\, s^2\rmd s; \quad {\cal N}\equiv n(\xi_B,\tau).
\end{equation}

Substituting Eq.~(\ref{Mad}) into Eq.~(\ref{feq1}) and separating the real and imaginary
parts, we arrive at a pair of equations:
\begin{eqnarray}
&&2\partial_\tau\chi+2\partial_\xi\chi\,\partial_\xi\theta+\chi\Delta_\xi\theta=0,
\label{eeq1}\\
&&2\partial_\tau\theta+(\partial_\xi\theta)^2-\frac{1}{\chi}\Delta_\xi\chi+W[\chi]=0;
\label{eeq2}\\
&&W[\chi]\equiv A\Phi_{\rm gr}+j_0\left(\chi\right)-1.
\label{W}
\end{eqnarray}

As shown in \cite{Naz2025}, the self-interaction $j_0(\chi)$
induces an alternation of phases: unstable for $\chi\in[a_{2n-1};a_{2n}]$
and stable for $\chi\in[a_{2n};a_{2n+1}]$ phases, where the sequence $\{a_n\}$
is defined by the zeros of $j_1(a_n)=0$, with $a_1=0$ and $a_{n-1}<a_n$.
Although infinitely many such phases exist, in practice we consider only two:
an unstable one for $\chi<a_2$, and a stable one for $a_2<\chi<a_3$.

Since the framework is developed to characterize the DM distribution, 
it must, akin to other self-interacting models, address the core-cusp 
problem at the origin~\cite{Lee96,Harko2011}. The regularity condition 
$\chi^\prime(0,\tau)=0$, dictated by spherical symmetry, effectively 
smooths the central density peak. However, evaluating the core value 
$\chi(0,\tau)$ within a fully nonlinear framework remains highly 
nontrivial; a detailed treatment of this issue is provided 
in~\cite{Naz2025,GN23}.

Besides, we initially assume the existence of a solution within
a finite spatial domain, \mbox{$\xi\in[0; \xi_B]$}, where the boundary value $\xi_B$ is
determined by the condition $j_0(\chi(\xi_B,\tau))=0$. For convenience, we
set $\chi(\xi_B,\tau)=\pi$, so that the interval $\xi\in[0; \xi_B]$
encompasses both phases. The correspondence between $\xi_B$ and dimensional $R$ 
is governed by the scale $r_0$. Meanwhile, the phase $\theta$ depends significantly
on the chosen physical regime.

In the case of a {\it coherent} BEC, the DM is characterized
by a fixed quantity -- a constant chemical potential
\begin{equation}\label{muu}
\widetilde{\mu}=\frac{\hbar^2}{mr_0^2}u,
\end{equation}
which is parametrized by a dimensionless $u<0$. Therefore,
the phase takes on the form:
\begin{equation}\label{th0}
\theta=-\frac{\widetilde{\mu}t}{\hbar}=-u\tau.
\end{equation}

In a spherical DM model, the constancy of its chemical potential
may be attributed to the uniform gravitational potential produced
inside the sphere by the isotropic outer shell of surrounding matter.

Substituting Eq.~(\ref{th0}) in Eq.~(\ref{eeq2}) leads to stationary GPP
equations with an effective parameter $\nu=1+2u-A\Phi_0$:
\begin{subequations}\label{eqqs}
\begin{align}
&(\Delta_\xi+\nu)\chi-A\Phi\chi-\sin{\chi}=0,
\label{eqq1}\\
&\Phi=\int_0^{\xi}\left(\frac{1}{s}-\frac{1}{\xi}\right)\chi^2(s)s^2\rmd s.
\label{eqq2}
\end{align}
\end{subequations}
Solutions and macroscopic properties of such a model have been studied
previously in \cite{Naz2025}, including applications to the dwarf
galaxy NGC 2366.

The case of a nonuniform phase, when $\partial_\xi\theta\not=0$,
corresponds to {\it decoherent} BEC DM.
The dephasing of parts of the condensate can be caused by an external
impact with the momentum transfer, which leads to the emergence of
a DM flow with a local speed
\begin{equation}\label{vdef}
v(\xi,\tau)=\partial_\xi\theta(\xi,\tau).
\end{equation}
The symmetry of the model eliminates the vortex contribution
to the velocity $v$, leaving us only the gradient part.

Multiplying Eq.~(\ref{eeq1}) by $\chi$ and taking into account the definition
of the particle number density $\eta=\chi^2$, one obtains the continuity
equation for a spherical model:
\begin{equation}\label{eeq11}
\partial_\tau\eta+\frac{1}{\xi^2}\partial_\xi\left(\xi^2\eta v\right)=0.
\end{equation}

On the other hand, by differentiating Eq.~(\ref{eeq2}) with respect to $\xi$,
we arrive at the Euler hydrodynamic equation:
\begin{equation}\label{eeq21}
\partial_\tau v+\partial_\xi\left[\frac{v^2}{2}
-\frac{1}{2\sqrt{\eta}}\Delta_\xi\sqrt{\eta}+\frac{1}{2}W[\sqrt{\eta}]\right]=0.
\end{equation}

In the next Section, we extend these equations by including terms that
account for the gravitational perturbations induced by baryonic matter.
By progressively incorporating such perturbations over time, the stationary
solutions of Eqs.~(\ref{eqqs}) serve as initial conditions. In Section~\ref{S4},
we return to the autonomous equations to investigate the quantum aspects
of DM excitations.

%% %% %% %% %% %% %% %% %% %% %% %% %% %% %% %% %% %% %% %% %% %% %% %% %% %% %% %%
\section{The impact of baryonic perturbations on DM}\label{S3}

\subsection{Clustering of baryonic matter. The case of dwarf NGC 2366}

To capture the qualitative effects of star formation and stellar feedback on
BEC DM under spherical symmetry, we intend to extract the central star/gas
cluster (overdensity) as a spherical shell near the galactic center,
independent of the overall shape of the galaxy.
We distribute baryonic matter (stars and gas) unevenly across the radius in
a spherical shell whose thickness is governed by the scale $d$, allowing $d$
to contract during collapse and grow during explosive events.
Although the processes that redistribute baryons are complex and omitted here,
our interest lies in the consequent changes to the gravitational potential and
the resulting response of DM.

We generalize the Plummer spherical profile of density~\cite{Plum}, originally
centered at the origin, to a baryonic cluster whose center is currently defined
by radius-vector $\breve{{\bf r}}$ by replacing ${\bf r}\to{\bf r}-\breve{{\bf r}}$.
To obtain a spherical shell model we then average this shifted profile over
all orientations, using $s\in[-1;1]$ (cosine of the angle between ${\bf r}$
and $\breve{{\bf r}}$), producing a normalized radial distribution that depends
only on the scalar radius $r$:
\begin{eqnarray}
\eta_{\rm b}(r,\breve{r},d)&=&\frac{3d^2}{2}\int_{-1}^{1}\frac{\rmd s}
{(r^2-2r\breve{r}s+\breve{r}^2+d^2)^{5/2}}
\nonumber\\
&=&\frac{d^2}{2r\breve{r}}\left[\frac{1}{[(r-\breve{r})^2+d^2]^{3/2}}-
\frac{1}{[(r+\breve{r})^2+d^2]^{3/2}}\right];
\label{n1}\\
&&\hspace*{-20mm}
\int_0^{\infty}\eta_{\rm b}(r,\breve{r},d)\,r^2\rmd r=1.
\end{eqnarray}
In particular, the Plummer profile is restored as
\begin{equation}
\lim\limits_{\breve{r}\to0}\eta_{\rm b}(r,\breve{r},d)=\frac{3d^2}{(r^2+d^2)^{5/2}}.
\end{equation}

Note that various aspects of shells in general relativity and astrophysics
are in~\cite{Isr66,EL77,Wh94}.

To account for realistic deviations from spherical symmetry, we can
generalize our approach by constructing an ellipsoidal baryon shell model.
Assuming that the baryonic cluster undergoes an axisymmetric ellipsoidal deformation,
characterized by the polar radius $R_d$ in the midplane and the scale height $h$
along the vertical axis, we introduce the effective core radius $\breve{r}=\sqrt{(R_d^2 + h^2)/2}$
and the eccentricity parameter $\varepsilon=(R_d^2 - h^2)/(R_d^2 + h^2)$, which
allows us to derive an analogue of the profile $\eta_{\rm b}(r,\breve{r},d)$.
Defining the spatial boundary as $R(\theta)=\breve{r}\sqrt{1-\varepsilon\cos(2\theta)}$,
the modified distribution reads
\begin{equation}\label{geta}
\eta_{\rm b}(r,\theta,\breve{r},d,\varepsilon)=\frac{3d^2}{4\pi}\int
\frac{\sin{\theta^\prime}\rmd\theta^\prime\rmd\phi^\prime}
{\left[r^2+R^2(\theta^\prime)
-2r\breve{r}\left(\sqrt{1-\varepsilon}\cos{\theta}\cos{\theta^\prime}
+\sqrt{1+\varepsilon}\sin{\theta}\sin{\theta^\prime}\cos{\phi^\prime}\right)+d^2\right]^{5/2}},
\end{equation}
where the integration domains $\phi^\prime\in[0;2\pi)$, $\theta^\prime\in[0;\pi]$ are used.

In the spherical limit $\varepsilon\to0$, the expression (\ref{geta}) reduces to
the isotropic shell profile (\ref{n1}). Given the complexity of analytically
calculating the integral (\ref{geta}), its properties, including the deviations
of the corresponding gravitational potential from sphericity in the clusters used,
are numerically analyzed below.

The shell profile (\ref{n1}) is asymmetric, with a peak at
\begin{equation}
r_{\rm p}\simeq\breve{r}\left[1-\frac{d^2}{3\breve{r}^2}
-\frac{d^4}{18\breve{r}^4}+O\left(\frac{d^5}{\breve{r}^5}\right)\right],
\end{equation}
when $d<\breve{r}$; and the profile decays
rapidly over the scale $d$ for the deviation $|r-r_{\rm p}|$.

The total baryon mass $M$ of such a cluster is distributed uniformly
over solid angle so that $\rmd^2 M/\rmd\Omega^2=M/(4\pi)$,
while the volumetric mass density is
\begin{equation}
4\pi\frac{\rmd^3 M}{r^2\,\rmd r\,\rmd\Omega^2}=M
\,\eta_{\rm b}(r,\breve{r},d).
\end{equation}

Estimating the integral
\begin{equation}
\int_{\breve{r}-d}^{\breve{r}+d}\eta_{\rm b}(r,\breve{r},d)\,r^2\rmd r=
\frac{\sqrt{2}}{2}-\frac{1}{8}\frac{d^3}{\breve{r}^3}
+O\left(\frac{d^5}{\breve{r}^5}\right),
\end{equation}
we conclude that the spherical layer $r\in[\breve{r}-d; \breve{r}+d]$
contains mass $\sim0.7M$ if $d\ll\breve{r}$. Then the mean mass density
of the layer is estimated as
\begin{equation}\label{avrho}
\overline{\rho}_{\rm b}\simeq
\frac{M}{16\pi} \frac{\sqrt{2}}{d\breve{r}^2}.
\end{equation}

Moreover, when dealing with observational data, we need to introduce
the surface density $\Sigma_0(R,\breve{r},d)$ associated with the spatial
distribution $\eta_{\rm b}(r,\breve{r},d)$. Analytically, the function
$\Sigma_0(R,\breve{r},d)$ is written in terms of elliptic integrals and
has a complex form in Appendix~\ref{ApA}. Then, by extracting the parameters
of each cluster using it, we can approximately reconstruct its spatial
distribution.

Here we turn to observational data on the distribution of gas (G) and
stars (S) in the dwarf galaxy NGC~2366 from \cite{Oh08}. We approximate
the surface density profiles using six clusters:
\begin{equation}
\Sigma(R)=\sum\limits_{i=1}^6 \frac{M_i}{4\pi}\,\Sigma_0(R,\breve{r}_i,d_i).
\end{equation}

\begin{table}[!t]
\begin{minipage}[b]{0.48\linewidth}
\centering
\caption{Parameters of the gas clusters.}
\begin{tabular}{|c|ccc|}
\hline
 cluster & $\breve{r}_i~[\text{kpc}]$  &  $d_i~[\text{kpc}]$ & $M_i~[10^6 M_\odot]$ \\
\hline
1 & 1.10 & 0.23 & 8.042 \\
2 & 1.66 & 0.51 & 38.20 \\
3 & 2.52 & 0.64 & 145.8 \\
4 & 3.76 & 0.84 & 202.1 \\
5 & 4.84 & 1.24 & 264.7 \\
6 & 6.57 & 1.91 & 189.1 \\
\hline
\end{tabular}
\label{tabG}
\end{minipage}
\hspace{0.3cm}
\begin{minipage}[b]{0.48\linewidth}
\centering
\caption{Parameters of the star clusters.}
\begin{tabular}{|c|ccc|}
\hline
 cluster & $\breve{r}_i~[\text{kpc}]$  &  $d_i~[\text{kpc}]$ & $M_i~[10^6 M_\odot]$ \\
\hline
1 & 0.46 & 0.24 & 12.32 \\
2 & 1.10 & 0.09 & 16.34 \\
3 & 1.45 & 1.01 & 13.07 \\
4 & 2.04 & 0.03 & 22.37 \\
5 & 3.17 & 1.42 & 138.7 \\
6 & 5.97 & 2.00 & 113.1 \\
\hline
\end{tabular}
\label{tabS}
\end{minipage}
\end{table}

\begin{figure}[tbp]
\centering
\captionsetup{width=0.9\textwidth}
\includegraphics[width=0.6\textwidth]{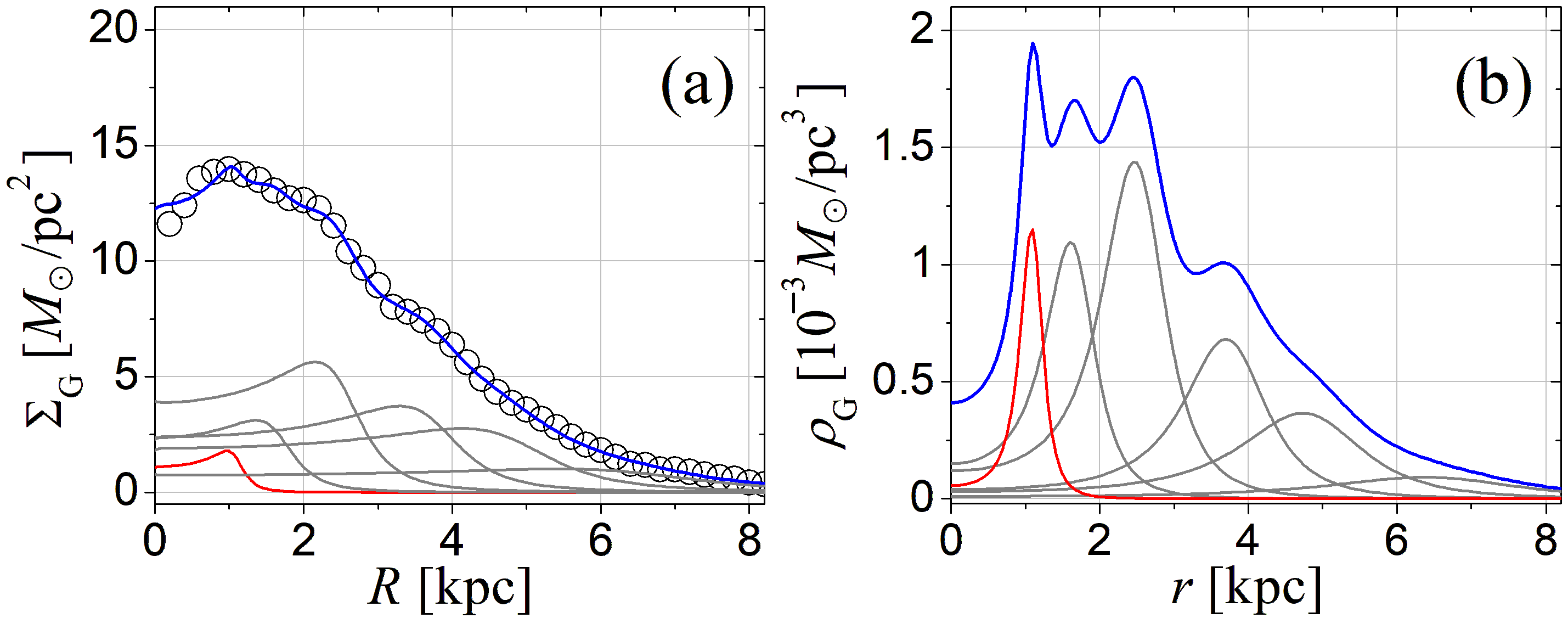} %% 8sm
%\vspace*{-1mm}
\caption{\label{gas}Interstellar gas distributions in NGC~2366.
The blue curves envelope cluster contributions, colored in gray
and red. Circles in panel (a) are the data from \cite{Oh08}.}
\end{figure}

\begin{figure}[!t]
\centering
\captionsetup{width=0.9\textwidth}
\includegraphics[width=0.6\textwidth]{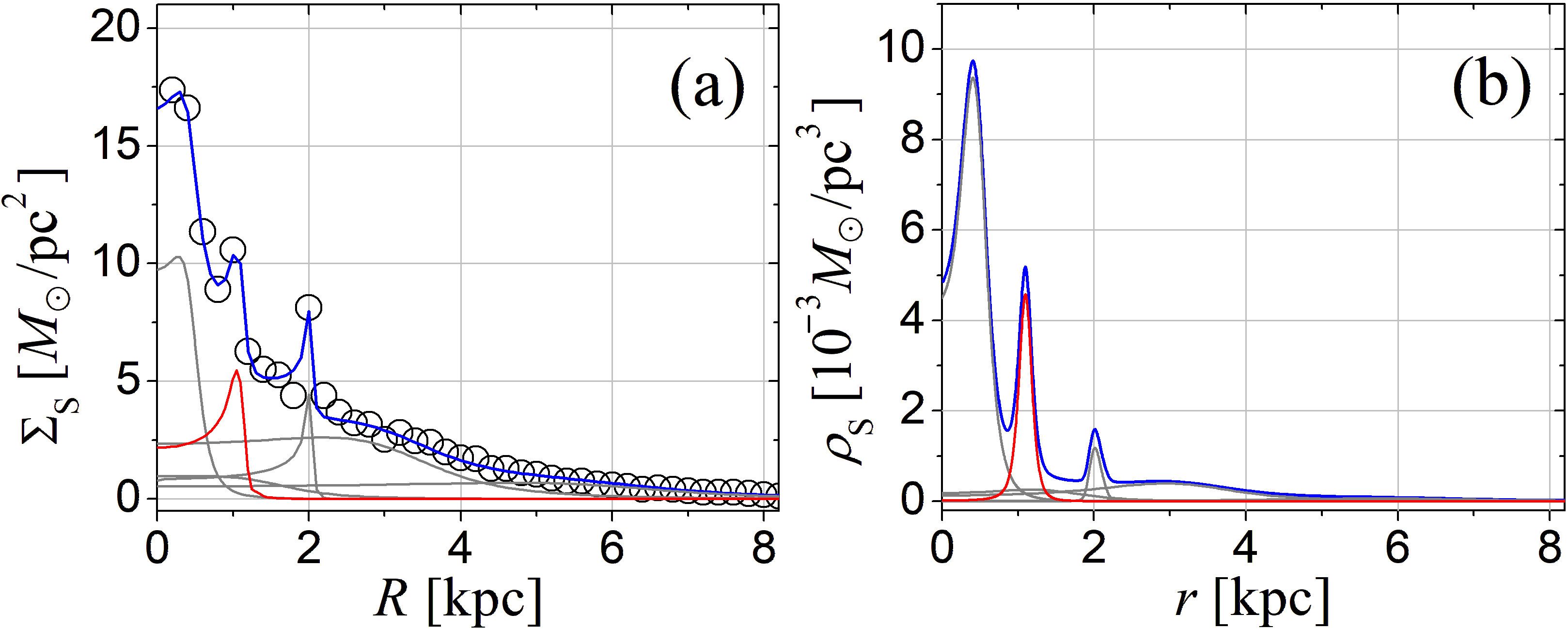} %% 8sm
%\vspace*{-1mm}
\caption{\label{stars}Stellar distributions in NGC~2366.
The blue curves envelope the cluster contributions,
colored in gray and red. Circles in panel (a) are the data
from \cite{Oh08}.}
\end{figure}

Cluster decompositions with the parameters from Tables~\ref{tabG}
and \ref{tabS} are shown in Figs.~\ref{gas}a and \ref{stars}a.
Reconstructing the model spatial distributions as
\begin{equation}
\rho(r)=\sum\limits_{i=1}^6 \frac{M_i}{4\pi}\,\eta_{\rm b}(r,\breve{r}_i,d_i),
\end{equation}
the obtained profiles are shown in Figs.~\ref{gas}b and \ref{stars}b.

Analyzing the cluster data, we note that gas cluster-1 is located at
$\breve{r}=1.1~\text{kpc}$, coinciding with the shell core position of
star cluster-2. Both clusters are colored red in Figs.~\ref{gas} and \ref{stars}.
This spatial overlap may suggest either a local accumulation of gas around
the stars of cluster-2 or gas ejection originating from them. Given
the relatively high gas mobility, we next consider how changes in the
width $d_1$ of its spatial distribution, arising in both scenarios, may
exert gravitational influence on the surrounding DM.

The presence of star cluster-1 near the galactic center suggests
an additional scenario: the possible inward approach of star cluster-2,
accompanied by a decrease in $\breve{r}_2$. 
Such a motion could be caused by the migration of stars
toward the galactic center, a process usually explained by dynamical
friction.

These baryonic clusters, which remain approximately spherical within
$r\leq1~\text{kpc}$, spatially overlap the dense phase of the DM core,
which will experience the strongest perturbations during the hypothetical
processes. According to the DM profile for NGC 2366 in \cite{Naz2025},
this phase is determined by the amplitude range
$a_2=4.493409<\chi<6.551148=\chi_0$ for $r<2.48~\text{kpc}$.

When modeling the processes, we nondimensionalize the quantities.
Since the chosen baryon distribution is a homogeneous function,
we simply have
\begin{equation}
\eta_{\rm b}(r,\breve{r},d)=r_0^{-3}\eta_{\rm b}(\xi,\breve{\xi},\ell),
\end{equation}
where
\begin{equation}
\xi=\frac{r}{r_0},\quad \breve{\xi}=\frac{\breve{r}}{r_0},\quad
\ell=\frac{d}{r_0}.
\end{equation}

Using $r_0=1.18~\text{kpc}$ and $m=0.1171\times10^{-22}~\text{eV}c^{-2}$
for DM in NGC 2366 from \cite{Naz2025}, we introduce the scales of velocity and time:
\begin{eqnarray}
&&v_0\equiv\frac{\hbar}{mr_0}\simeq1.387451\times10^5~\text{m}/\text{s},
\label{vel0}\\
&&t_0\equiv\frac{mr_0^2}{\hbar}\simeq2.624308\times10^{14}~\text{s}\simeq8.316\times10^6~\text{yr}.
\end{eqnarray}
These scales are used in the rest of the article.

%% %% %% %% %% %% %% %% %% %% %% %% %% %% %% %% %% %% %% %% %% %% %% %% %% %% %% %%
\subsection{Partial gravitational potentials}

According to Gauss theorem, the gravitational potential of a {\it thin} shell
remains constant inside the shell, but outside it obeys Newton's inverse-radius law.
Thickening the shell to a small finite width only softens local features and does not
change the large-scale trend, making the detailed shape of
$\eta_{\rm b}(\xi,\breve{\xi},\ell)$ irrelevant. The use of the modified Plummer
profile merely simplifies the analytical description.

Based on these ideas, the Poisson equation 
$\Delta_\xi\Phi_{\rm b}=\eta_{\rm b}(\xi,\breve{\xi},\ell)$, with $\Delta_\xi$
the radial part of the Laplacian, yields the potential:
\begin{eqnarray}
&&\hspace*{-3mm}
\Phi_{\rm b}(\xi,\breve{\xi},\ell)=-\frac{\sqrt{(\xi+\breve{\xi})^2+\ell^2}
-\sqrt{(\xi-\breve{\xi})^2+\ell^2}}{2\xi \breve{\xi}};\\
&&\hspace*{-3mm}
\Phi_{\rm b}(\xi,0,\ell)=-\frac{1}{\sqrt{\xi^2+\ell^2}}.
\end{eqnarray}
The value $\Phi_{\rm b}(0,\breve{\xi},\ell)$ is obtained from the last expression
by interchanging $\xi$ and $\breve{\xi}$.

\begin{figure}[htbp]
\centering
\captionsetup{width=0.9\textwidth}
\includegraphics[width=0.7\textwidth]{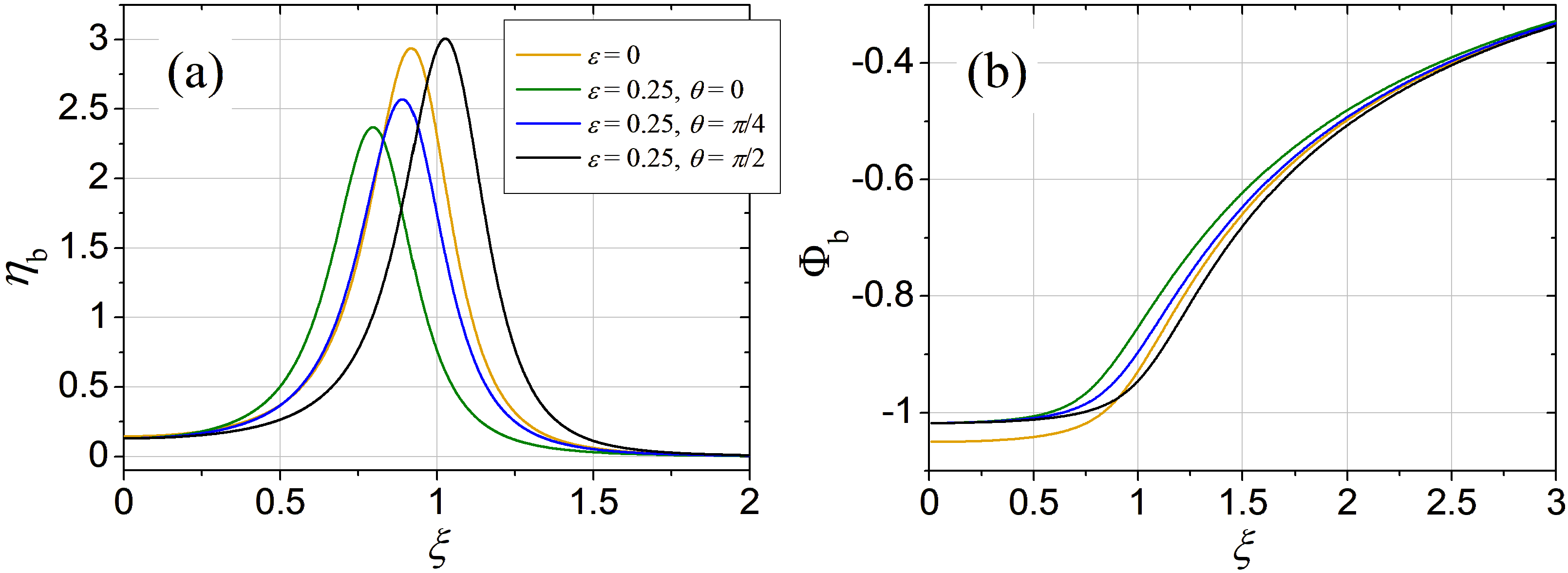}
\caption{\label{ellp}Radial profiles of (a) the baryonic density
distribution $\eta_{\rm b}(\xi, \theta)$ and (b) the corresponding gravitational 
potential $\Phi_{\rm b}(\xi, \theta)$ for gas cluster-1 ($\breve{\xi}=0.932$, $\ell=0.195$).
The baseline spherically averaged profile ($\varepsilon=0$) is represented
by the orange curve, while the exact ellipsoidal configurations ($\varepsilon=0.25$)
are evaluated at observer polar angles $\theta=0$ (green), $\theta=\pi/4$ (blue),
and $\theta=\pi/2$ (black).}
\end{figure}

On the other hand, determining the exact gravitational potential $\Phi_{\rm b}(\xi,\theta)$
generated by an ellipsoidal configuration (\ref{geta}) requires the use of numerical approach.
Performing the calculations for gas cluster-1 ($\breve{\xi}=0.932$, $\ell=0.195$), we observe
(see Fig.~\ref{ellp}) that in the inner region of the core ($\xi\to0$), the potential
profiles smoothly converge to a uniform plateau. This explicitly demonstrates that
non-spherical multipole corrections vanish in the dynamics and validate the applicability
of the symmetric shell ansatz.

The total gravitational potential created by all baryonic clusters is equal to
\begin{equation}
V_{\rm b}(\xi)=\sum_{k\in\text{all clusters}} A_{\rm b}^{(k)}
\Phi_{\rm b}\left(\xi,\frac{\breve{r}_k}{r_0},\frac{d_k}{r_0}\right),
\end{equation}
where $A_{\rm b}^{(k)}$ is the dimensionless coupling for baryonic cluster-$k$
with the total mass $M_k$:
\begin{equation}
A_{\rm b}^{(k)}=2\frac{G M_k m^2 r_0}{\hbar^2}.
\end{equation}
Substituting the parameters of NGC 2366's DM, the overall coefficient is estimated as
\begin{equation}\label{Ab}
A_{\rm b}\simeq3.80\times10^{-4}\,\left[\frac{M}{10^6M_\odot}\right].
\end{equation}

Due to its smallness, the initial gravitational background inherent to baryonic
matter perturbations in Eqs.~(\ref{eqqs}) is absorbed by the chemical potential $u$
from Eq.~(\ref{muu}) under DM dominance, as
\begin{equation}\label{uu}
2u=2u_0+\langle V_{\rm b}\rangle.
\end{equation}
Here the second term represents the quantum mean of the gravitational
potential and corresponds to the first correction of perturbation theory,
the mean field:
\begin{equation}\label{1Vc}
\langle V_{\rm b}\rangle=\langle\phi|V_{\rm b}|\phi\rangle
=\frac{1}{{\cal N}}\int_0^{\xi_B}V_{\rm b}(s)\,\chi^2(s)\,s^2\rmd s,
\end{equation}
see Eqs.~(\ref{Mad}) and (\ref{Nt}) for using unnormalized $\phi$.

By reproducing the rotation curve for NGC~2366 in \cite{Naz2025},
the equilibrium DM distribution $\eta(\xi)\equiv\chi^2(\xi)$ in
Fig.~\ref{ceta}a below is obtained for the constant chemical
potential $2u=-1.5867$. Using this fact and substituting the cluster
parameters from Table~\ref{tabG} and \ref{tabS}, we calculate that
$\langle V_{\rm b}\rangle=-0.0926$ and, accordingly, $2u_0=-1.4941$.
This confirms the approximation (\ref{uu}) and may indicate an external
source for the main contribution, $2u_0$.

Although the potential $V_{\rm b}$ includes contributions from all feasible
clusters, we intend to analyze the impact caused by restructuring a single
cluster, avoiding the DM condensate depletion. By synchronously redistributing
the baryons in the selected cluster, we obtain a new gravitational potential:
\begin{equation}
V^\prime_{\rm b}=V_{\rm b}+A_{\rm b}\Delta\Phi_{\rm b},
\end{equation}
where $\Delta\Phi_{\rm b}$ denotes the change in the gravitational
potential, defined as the final potential minus the initial potential.

We assume mass conservation for the cluster: its total mass $M$
remains fixed while the cluster's profile/position may vary.
Given Eq.~(\ref{uu}), the perturbed baryonic contribution acting on the DM,
to the same order of approximation, becomes
\begin{equation}\label{pv}
\langle V_{\rm b}\rangle\to\langle V_{\rm b}\rangle+A_{\rm b}\Delta\Phi_{\rm b}.
\end{equation}
Such changes occur adiabatically:
\begin{equation}
\left|\frac{\Delta\Phi_{\rm b}}{\Phi_{\rm b}}\right|\ll1.
\end{equation}

Since $A_{\rm b}|\Delta\Phi_{\rm b}|\ll|V_{\rm b}|$, the perturbation produces
only a negligible change in the DM density profile. The condensate, however,
does acquire a spatially varying phase. This induced phase inhomogeneity
drives decoherence of the condensate and generates macroscopic flows.

With small changes in the geometric parameters of the baryon distribution,
we can write:
\begin{equation}\label{dPhi}
\delta\Phi_{\rm b}(\xi,\breve{\xi},\ell)=\Phi_{\ell}(\xi,\breve{\xi},\ell)\delta\ell
+\Phi_{\xi}(\xi,\breve{\xi},\ell)\delta\breve{\xi}.
\end{equation}
where $\Phi_{\ell}\equiv\partial\Phi_{\rm b}/\partial\ell$
and $\Phi_{\xi}\equiv\partial\Phi_{\rm b}/\partial\breve{\xi}$.

\begin{figure}[!t]
\centering
\captionsetup{width=0.9\textwidth}
\includegraphics[width=0.6\textwidth]{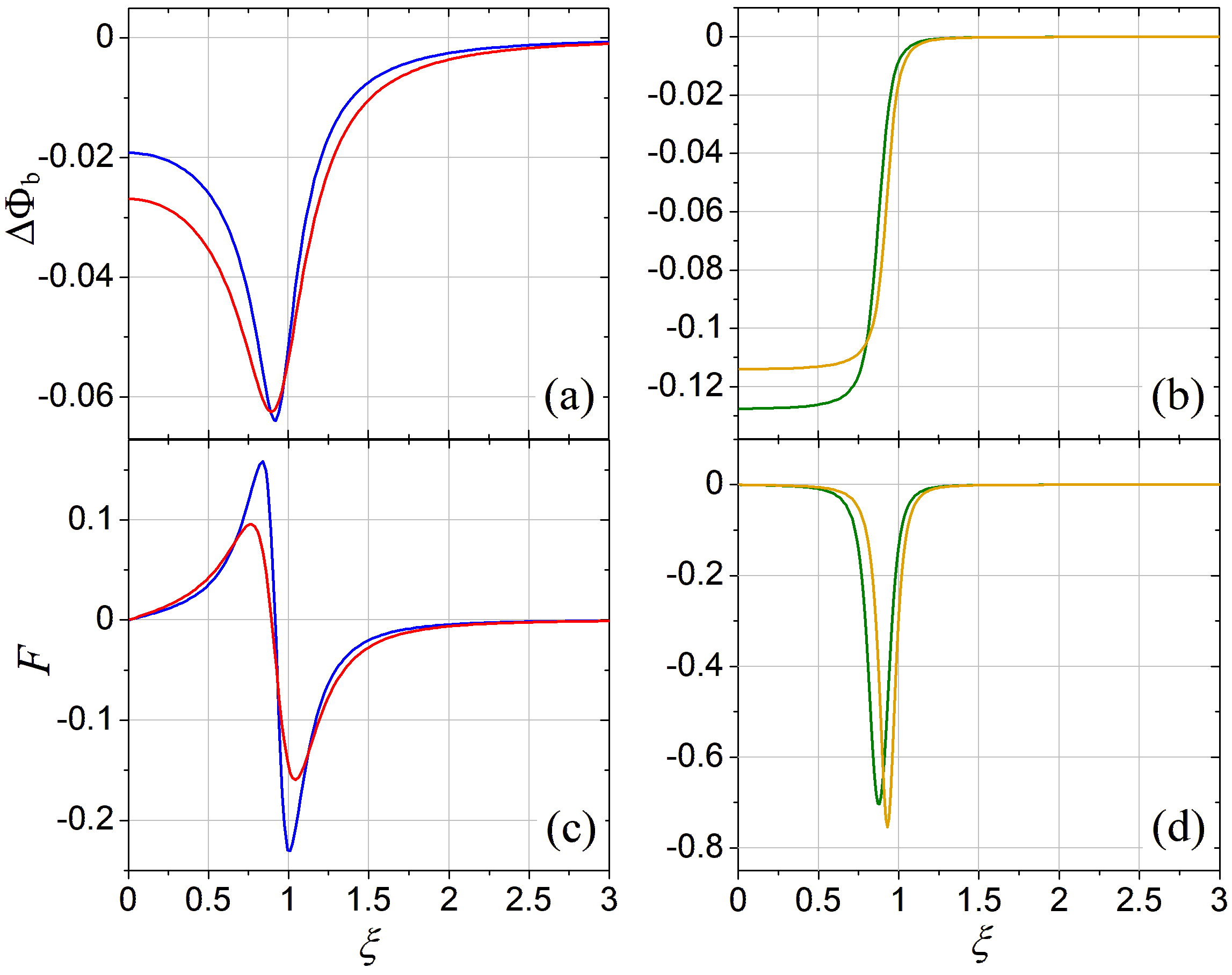} %% 8sm
%\vspace*{-1mm}
\caption{\label{pots}Profiles of the baryon gravitational potential
variations and corresponding forces for NGC 2366 scenarios. 
Top row: Gravitational potential differences $\Delta\Phi_{\rm b}$
(blue and green lines) and linear variations $\delta\Phi_{\rm b}$
(red and orange lines) induced by (a) gas shell narrowing
(width $\ell$ compression from $0.195$ to $0.076$ at fixed
core radius $\breve{\xi}=0.932$), and (b) leftward displacement
of the cluster ($\breve{\xi}$ shift by $-0.1$ at fixed $\ell=0.076$). 
Bottom row: Corresponding tidal forces $F$ derived via 
$-\partial_\xi\Delta\Phi_{\rm b}$ and $-\partial_\xi\delta\Phi_{\rm b}$, 
maintaining matching color codings for each scenario.}
\end{figure}

Formula (\ref{dPhi}) indicates two types of possible processes associated
either with the matter redistribution in the shell or with a displacement
of the entire shell. Moreover, it provides a basis for variable separation,
allowing the potential change to be expressed as
$\delta\Phi_{\rm b}=\Phi_x(\xi)\delta x(\tau)$.

We easily find that
\begin{eqnarray}
&&\hspace*{-4mm}
\Phi_{\ell}(\xi,\breve{\xi},\ell)=\frac{\ell}{2\breve{\xi} \xi}\left[\frac{1}{J_{-}(\xi,\breve{\xi},\ell)}
-\frac{1}{J_{+}(\xi,\breve{\xi},\ell)}\right],
\label{Phil}\\
&&\hspace*{-4mm}
\Phi_{\xi}(\xi,\breve{\xi},\ell)=-\frac{1}{2\breve{\xi}^2 \xi}\left[\frac{\xi^2-\xi\breve{\xi}+\ell^2}{J_{-}(\xi,\breve{\xi},\ell)}
-\frac{\xi^2+\xi\breve{\xi}+\ell^2}{J_{+}(\xi,\breve{\xi},\ell)}\right],
\label{Phix}
\end{eqnarray}
where
\begin{equation}
J_{\pm}(\xi,\breve{\xi},\ell)\equiv\sqrt{(\xi\pm\breve{\xi})^2+\ell^2}.
\end{equation}

Assuming that the time dependence is concentrated in $\delta x(\tau)$,
we adopt the relaxation-time approximation for all processes, thereby
unifying our approach:
\begin{equation}
\frac{\rmd^2\delta x}{\rmd\tau^2}+\frac{1}{\tau_{ch}}\frac{\rmd\delta x}{\rmd\tau}=0,
\quad \delta x(0)=0, \quad \delta x(\tau)\xrightarrow{\tau\rightarrow\infty}\delta x_\infty,
\end{equation}
where $\tau_{ch}$ is the characteristic relaxation time;
and $\delta x_\infty$ is the asymptotic (final) value.

It means that the velocity $\delta{\dot x}(\tau)$ decays over time,
driven either by a counteracting force or by saturation effects. Such
a solution is given by the expression:
\begin{equation}\label{rta}
\delta x(\tau)=\delta x_\infty\left(1-\rme^{-\gamma\tau}\right),\quad
\gamma\equiv\frac{1}{\tau_{ch}}.
\end{equation}

It should be noted that the relaxation-time ansatz (\ref{rta})
serves as a collective kinematic parametrization for a smooth transition
between two baryonic states. Physically, due to quantum pressure,
the ultralight DM BEC acts as a low-pass filter,
rendering it insensitive to small-scale baryonic fluctuations that lack
significant gradients. Therefore, a smooth exponential transition is a physically
sufficient approximation for the DM forced response. Although a full treatment
of stellar orbits would require integrating the equations of motion with
angular momentum loss, the initial dynamical friction rate derived in
the next Section~3.3 strictly grounds the realistic adiabatic timescale
$\tau_{ch}$ of this transition.

Thus, $\delta\Phi_{\rm b}=\Phi_{\ell}(\xi,\breve{\xi},\ell)\delta\ell$
provides a linear approximation to the difference
$\Delta\Phi_{\rm b}=\Phi_{\rm b}(\xi,\breve{\xi},\ell+\delta\ell)-
\Phi_{\rm b}(\xi,\breve{\xi},\ell)$, which arises from a temporal
variation of $\delta\ell(\tau)$. The case of $\breve{\xi}=0.932$,
$\ell=0.195$, and the final $\delta\ell_\infty=-0.119$, used below,
is shown in Fig.~\ref{pots}a.
A change in gravitational potential results in the emergence of a force,
determined by
\begin{equation}
-\partial_\xi\Phi_\ell
=\frac{\ell}{2\breve{\xi} \xi^2}\left(\frac{1}{J_{-}}
-\frac{1}{J_{+}}\right)
+\frac{\ell}{2\breve{\xi} \xi}\left(\frac{\xi-\breve{\xi}}{J^3_{-}}
-\frac{\xi+\breve{\xi}}{J^3_{+}}\right).
\end{equation}
The dependence of this force on time is also given by the factor $\delta\ell(\tau)$
for some $\tau_{ch}$.

Under the specified parameters, the collapse-induced compression of baryonic matter
from $\ell=0.195$ to $\ell=0.076$ yields the force profile presented in Fig.~\ref{pots}c.
Conversely, by reversing the force direction and changing its sign, one can evaluate
the force acting on DM, for instance, during expansion or explosion. It is essential,
however, to account for the varying influences that arise from different process rates
$\gamma$.

Another special case, $\delta\Phi_{\rm b}=\Phi_{\xi}(\xi,\breve{\xi},\ell)\delta\breve{\xi}$,
corresponds to a potential change
$\Delta\Phi_{\rm b}=\Phi_{\rm b}(\xi,\breve{\xi}+\delta\breve{\xi},\ell)-
\Phi_{\rm b}(\xi,\breve{\xi},\ell)$, arising from a cluster shift driven by
the time-dependent variation $\delta\breve{\xi}(\tau)$. For $\breve{\xi}=0.932$,
$\ell=0.076$, and the final value $\delta\breve{\xi}_\infty=-0.1$,
the resulting profiles are shown in Fig.~\ref{pots}b. The corresponding force can
then be estimated using the expression:
\begin{equation}
-\partial_\xi\Phi_\xi
=-\frac{\ell^2}{2\breve{\xi}^2 \xi^2}\left(\frac{1}{J_{-}}
-\frac{1}{J_{+}}\right)
+\frac{\ell^2}{2\breve{\xi} \xi}\left(\frac{1}{J^3_{-}}
+\frac{1}{J^3_{+}}\right).
\end{equation}

Although the contraction of an entire cluster toward the galactic center
occurs gradually, Fig.~\ref{pots}d demonstrates that the exerted force is
substantial. We therefore turn to examining the impact of cluster migration
on DM.

%% %% %% %% %% %% %% %% %% %% %% %% %% %% %% %% %% %% %% %% %% %% %% %% %% %% %% %%
\subsection{Migration rate of a star cluster under dynamical friction}

Although a rough estimate of $\tau_{ch}$ or $\gamma$ is sufficient, in principle,
to account for the gravitational influence of star-cluster migration on DM, we now
examine the migration process in greater detail. In our modeling, we focus on
the extracted shell-like star cluster-2, embedded within the wider gas cluster-1
of NGC~2366. Under these conditions, studying the dynamical friction mechanism,
which has traditionally been analyzed for a pointlike perturber after
Chandrasekhar's work~\cite{Chandra43}, becomes particularly relevant.

Let us isolate from the spherical star cluster-2 a toroidal tube (ring) with
a cross-sectional diameter of $2d_{\rm sc}$, lying on the equator with $r=\breve{r}$.
According to Eq.~(\ref{avrho}), the mean mass density of the star cluster-2 
is $\overline{\rho}_{\rm sc}\simeq2.877\times10^{-22}~\text{kg}/\text{m}^3$
at $M_{\rm sc}\simeq16.34\times10^6M_\odot$, and the tube's volume is
$\upsilon_{\rm tube}=2\pi^2\breve{r}d_{\rm sc}^2\simeq0.176~\text{kpc}^3$,
which gives us its mass 
$\mu=\overline{\rho}_{\rm sc}\upsilon_{\rm tube}\simeq7.42\times10^5M_\odot$.

Then the spherical cluster is covered by a number of
$N=M_{\rm sc}/\mu\simeq22.01$ tubes, each of which has a moment of inertia
$I=\mu\breve{r}^2$ at $d_{\rm sc}\ll\breve{r}$ and its own axis
of rotation with the same angular frequency $\Omega(\breve{r})$
determined by the rotation curve of the galaxy.

The total kinetic energy of the star cluster is
\begin{equation}
K_{\rm sc}=N\frac{I\Omega^2(\breve{r})}{2}=\frac{M_{\rm sc}V^2_t(\breve{r})}{2},
\end{equation}
where the tangential velocity $V_t(\breve{r})=\breve{r}\Omega(\breve{r})$.
Here, $V_t\simeq22~\text{km}/\text{s}$, as inferred from
the rotation curve in \cite{Naz2025}, which allows us to consider the effect
in the supersonic regime.

Thus, for simplicity, we densely cover a sphere of radius $r=\breve{r}$ with
stars rotating along circular equatorial orbits and collectively producing
the gravitational potential of a sphere:
\begin{equation}
W_{\rm sc}(r)=-\frac{GM_{\rm sc}}{\breve{r}}\,H(\breve{r}-r)
-\frac{GM_{\rm sc}}{r}\,H(r-\breve{r}),
\end{equation}
where the Heaviside function $H(r)$ is used.

On the other hand, we take into account the spatial distribution of
the mass $\rho_{\rm gc}(r,\breve{r},d_{\rm gc})$ in gas cluster-1,
which generates the gravitational potential
$GM_{\rm gc}\Phi(r,\breve{r},d_{\rm gc})$.

The change in the kinetic energy of the star cluster over time is written as
\begin{equation}
\dot K_{\rm sc}=M_{\rm sc}V_t{\dot V}_t
=-\frac{2\alpha}{t_{\rm df}}K_{\rm sc},
\end{equation}
where we define the rate and the slope:
\begin{equation}
\frac{1}{t_{\rm df}}\equiv-\frac{\dot{\breve{r}}}{\breve{r}}>0,\qquad
\alpha=\left.\frac{\rmd\ln{V_t(r)}}{\rmd\ln{r}}\right|_{r=\breve{r}}.
\end{equation}
We obtain that $\alpha\simeq0.94$ from the data in \cite{Naz2025}.

Therefore, one finds that
\begin{equation}
\frac{\alpha}{t_{\rm df}}=\left|\frac{\dot K_{\rm sc}}{2K_{\rm sc}}\right|=\left|\frac{{\dot V_t}}{V_t}\right|.
\end{equation}

Assuming that the change in the kinetic energy of the star cluster
is caused by a change in the gravitational potential of the gas wake
$W_{\rm w}$, we set
\begin{equation}
\left|\frac{\dot K_{\rm sc}}{2K_{\rm sc}}\right|=\left|\frac{\dot W_{\rm w}}{V^2_t}\right|.
\end{equation}

To find $\dot W_{\rm w}$, we allow a redistribution of the gas mass
driven by gravitational focusing near the surface $r=\breve{r}$:
\begin{equation}\label{difm}
\dot m_{\rm gc}(r,\breve{r},d_{\rm gc})=u(r)
\frac{\partial}{\partial d_{\rm gc}}m_{\rm gc}(r,\breve{r},d_{\rm gc}),
\end{equation}
where $u(r)$ is a radial velocity of cluster compression;
the mass under a sphere of radius $r$
\begin{equation}
m_{\rm gc}(r,\breve{r},d_{\rm gc})=4\pi\int_0^r
\rho_{\rm gc}(s,\breve{r},d_{\rm gc})\,s^2\rmd s,
\quad
m_{\rm gc}(\infty,\breve{r},d_{\rm gc})=M_{\rm gc}.
\end{equation}

Differentiating Eq.~(\ref{difm}) with respect to $r$, we arrive at
the continuity equation:
\begin{eqnarray}
&&\partial_t \rho_{\rm gc}+\frac{1}{r^2}\partial_r(r^2\rho_{\rm gc}v)=0,
\\
&&
v=-\frac{u(r)}{\rho_{\rm gc}(r,\breve{r},d_{\rm gc})}
\frac{\partial}{\partial d_{\rm gc}}\frac{m_{\rm gc}(r,\breve{r},d_{\rm gc})}{r^2}.
\end{eqnarray}

The time derivative of the excess gravitational force $F_{\rm w}$
per unit probe mass acting on the stars at $t=0$ is equal to
\begin{equation}
\dot F_{\rm w}=Gu(r)\frac{\partial}{\partial d_{\rm gc}}\frac{m_{\rm gc}(r,\breve{r},d_{\rm gc})}{r^2}
=-GM_{\rm gc}u(r)\frac{\partial^2}{\partial r\,\partial d_{\rm gc}}
\Phi(r,\breve{r},d_{\rm gc}).
\end{equation}

By relating the specific force $F_{\rm w}$ to $W_{\rm w}$, one has
\begin{equation}
\dot W_{\rm w}=-\int_0^\infty \dot F_{\rm w}\,\rmd r=-GM_{\rm gc}
\int_0^\infty \frac{\partial}{\partial d_{\rm gc}}
\Phi(r,\breve{r},d_{\rm gc})\,\rmd u(r),
\end{equation}
where we neglect the boundary terms by performing integration by parts.

Having defined the infinitesimal velocity $\rmd u(r)=-\partial_rW_{\rm sc}\rmd t$
according to the Newton equation, we can write that
\begin{equation}
\dot W_{\rm w}=-GM_{\rm gc}\int_{0}^{\infty}\rmd T(r)
\left(-\frac{\rmd W_{\rm sc}(r)}{\rmd r}\right) \,
\frac{\partial}{\partial d_{\rm gc}}\Phi(r,\breve{r},d_{\rm gc})
\end{equation}
where $T(r)=2\pi r/V_t(r)\simeq2\pi r/V_t(\breve{r})$ and
$\rmd T(r)\simeq2\pi\rmd r/V_t(\breve{r})$, since the integrand
is sharply peaked within a narrow radial interval.

Computing the integral by using Eq.~(\ref{Phil}) and dimensionless
variables, one gets
\begin{eqnarray}
&&\int_{\breve{\xi}}^{\infty}
\Phi_{\ell}(\xi,\breve{\xi},\ell)\,\frac{\rmd\xi}{\xi^2}
=\frac{x}{4\breve{\xi}^3}\left[\frac{2-x^2}{(1+x^2)^{5/2}}\,a(x)+\frac{b(x)}{(1+x^2)^2}\right];
\\
&&a(x)\equiv 2\,{\rm arctanh}\,\frac{1}{\sqrt{1+x^2}}
-{\rm arctanh}\,\frac{2+x^2}
{\sqrt{1+x^2}\sqrt{4+x^2}}
+{\rm arctanh}\,\frac{x}{\sqrt{1+x^2}},
\nonumber\\
&&b(x)\equiv x^3+4x-6+(2-x^2)\sqrt{4+x^2},
\nonumber
\end{eqnarray}
where $a(x)/\ln{x}\to-1$ and $b(x)/\ln{x}\to0$ at $x\equiv\ell/\breve{\xi}=d_{\rm gc}/\breve{r}\to0$.

In the limit $\breve{\xi}\gg1$ and $\ell\ll\breve{\xi}$, evaluating
this Coulomb-type integral yields
\begin{equation}
\int_{\breve{\xi}}^{\infty}
\Phi_{\ell}(\xi,\breve{\xi},\ell)\,\frac{\rmd\xi}{\xi^2}\simeq
\frac{\ell}{2\breve{\xi}^4}\ln{\frac{\breve{\xi}}{\ell}}.
\end{equation}

Using the mean gas density (\ref{avrho}) and combining
the formulas, we derive
\begin{equation}
\dot W_{\rm w}=16\sqrt{2}\pi^2 \frac{G^2M_{\rm sc}\overline{\rho}_{\rm gc}}{V_t}
\left(\frac{d_{\rm gc}}{\breve{r}}\right)^2\ln{\frac{\breve{r}}{d_{\rm gc}}}.
\end{equation}

Therefore, the initial rate of dynamical friction is given by
\begin{equation}\label{tdf1}
\frac{1}{t_{\rm df}}\simeq16\sqrt{2}\pi^2 \frac{G^2M_{\rm sc}\overline{\rho}_{\rm gc}}{\alpha V^3_t}
\left(\frac{d_{\rm gc}}{\breve{r}}\right)^2\ln{\frac{\breve{r}}{d_{\rm gc}}}.
\end{equation}
This expression differs from many other relaxation time estimates
for linear \cite{Chandra43,Rud71,Reph80,Ostr} and circular \cite{Tre84,Kim09,Lee13,Des22}
motions of perturber even in the supersonic regime, in a geometrical
factor that indicates a peculiarity of our two-spherical-shell model.
When the gas environment is replaced by DM from different
models~\cite{Lan20,Buh23,Lee25,Gor25,BEK19}, differences are also revealed.
However, the scale of dynamic friction for stars in a gaseous medium may be
compared for different geometries.

Thus, substituting the cluster parameters and
$\overline{\rho}_{\rm gc}\simeq5.5\times10^{-23}~\text{kg}\,\text{m}^{-3}$,
one has
\begin{equation}\label{tdf}
\frac{1}{t_{\rm df}}\simeq1.225\times10^{-17}~\text{s}^{-1},\qquad
\gamma=\frac{1}{\tau_{ch}}=\frac{t_0}{t_{\rm df}}\simeq0.0032.
\end{equation}
We intend to insert these values into the hydrodynamic framework of DM evolution,
assuming they remain constant as the star cluster shifts from
$\breve{r}=1.1~\text{kpc}$ by $0.1r_0$. The dynamics of the cluster itself
are omitted here.

%% %% %% %% %% %% %% %% %% %% %% %% %% %% %% %% %% %% %% %% %% %% %% %% %% %% %% %%
\subsection{Hydrodynamic DM evolution induced by three processes}

Given that the mass of an individual baryonic cluster is hierarchically smaller
relative to both the total baryonic inventory and the dominant DM halo mass,
the underlying DM density profile remains structurally unchanged in all three
scenarios proposed above. For this reason, the hydrodynamic evolution of 
DM perturbations driven by baryonic processes is naturally described by
linearized equations.

Therefore, for a relatively small magnitude of the velocity $v(\xi,\tau)=\delta v(\xi,\tau)$,
the quadratic term $v^2$ in Eq.~(\ref{eeq21}) can be omitted.
The variation of the DM density, $\delta\eta(\xi,\tau)$, relative to the
stationary distribution $\eta(\xi)$ that satisfies Eqs.~(\ref{eqqs}),
produces the following modifications in the interaction term and quantum
pressure:
\begin{eqnarray}
&&
W[\sqrt{\eta+\delta\eta}]=W[\sqrt{\eta}]+A\Delta^{-1}_\xi \delta\eta+c^2_s\frac{\delta\eta}{\eta}+O(\delta\eta^2),
\\
&&
\frac{\Delta_\xi\sqrt{\eta+\delta\eta}}{\sqrt{\eta+\delta\eta}}=
\frac{\Delta_\xi\sqrt{\eta}}{\sqrt{\eta}}+\frac{1}{2}
\widetilde{\Delta}_\xi\frac{\delta\eta}{\eta}+O(\delta\eta^2),
\end{eqnarray}
where the hydrodynamic Laplace operator $\widetilde{\Delta}_\xi$,
the square of the sound speed $c^2_s$, and the self-gravity potential
variation are
\begin{eqnarray}
&&
\widetilde{\Delta}_\xi\equiv\frac{1}{\xi^2\eta(\xi)}\partial_\xi
\xi^2\eta(\xi)\partial_\xi=\Delta_\xi+\frac{\partial_\xi\eta(\xi)}{\eta(\xi)}\partial_\xi,
\label{hDel}\\
&&
c^2_s\equiv\eta\frac{\rmd j_0\left(\sqrt{\eta}\right)}{\rmd\eta}=-\frac{\sqrt{\eta}}{2} j_1\left(\sqrt{\eta}\right),
\label{sss}\\
&&
\Delta^{-1}_\xi \delta\eta(\xi,\tau)=\int_0^{\xi}\left(\frac{1}{s}-\frac{1}{\xi}\right)
\delta\eta(s,\tau)\,s^2\rmd s+C(\tau).
\end{eqnarray}

The integral
\begin{equation}
C(\tau)=-\int_0^{\xi_B}\delta\eta(s,\tau)\,s\rmd s
\end{equation}
is irrelevant in the hydrodynamic equations, but it does
contribute to the phase $\theta$.

Since the DM density $\eta(\xi)$ is a solution to Eqs.~(\ref{eqqs}),
one has that
\begin{equation}
-\frac{\Delta_\xi\sqrt{\eta}}{\sqrt{\eta}}+W[\sqrt{\eta}]=2u,
\end{equation}
where $u$ is the constant chemical potential.

Given Eqs.~(\ref{uu}) and (\ref{pv}), necessary to account for the
redistribution of baryons, we need to subtract the gravitational
potential of the initial configuration from the chemical potential,
and then add the potential of the new configuration. To an
infinitesimal approximation, this yields the following replacement:
\begin{equation}
2u\to2u+A_{\rm b}\delta\Phi_{\rm b}(\xi,\tau),
\end{equation}
writing the time-dependent potential difference as
\begin{equation}
\delta\Phi_{\rm b}(\xi,\tau)=\Phi_{\ell}(\xi,\breve{\xi},\ell)\,\delta\ell(\tau)
+\Phi_{\xi}(\xi,\breve{\xi},\ell)\,\delta\breve{\xi}(\tau).
\end{equation}

We assume that the underlying equilibrium
density $\eta(\xi)$ remains unchanged in all our
scenarios. Moreover, we impose the same initial conditions:
\begin{equation}
\delta\eta(\xi,0)=0,\quad v(\xi,0)=0.
\end{equation}

Thus, the set of hydrodynamic equations reads
%
%\begin{eqnarray}
\begin{subequations}\label{HEqs}
\begin{align}
&\partial_\tau\delta\eta(\xi,\tau)+\frac{1}{\xi^2}
\partial_\xi\left[\xi^2 \eta(\xi) v(\xi,\tau)\right]=0,
\label{hyd1}\\
&
2\partial_\tau v(\xi,\tau)+\partial_\xi\mathbb{V}(\xi,\tau)=0,
\label{hyd2}
\end{align}
\end{subequations}
%\end{eqnarray}
%
where we have introduced auxiliary potential:
\begin{equation}\label{Vp}
\mathbb{V}(\xi,\tau)=A_{\rm b}\delta\Phi_{\rm b}(\xi,\tau)
+A\Delta^{-1}_\xi \delta\eta(\xi,\tau)
+\left[c^2_s(\xi)-\frac{\widetilde{\Delta}_\xi}{2}\right]
\frac{\delta\eta(\xi,\tau)}{\eta(\xi)}.
\end{equation}

Furthermore, we impose the boundary conditions:
%
%\begin{eqnarray}
\begin{subequations}
\begin{align}
&\delta\eta^\prime(0,\tau)=0,&\quad &v(0,\tau)=0,
\label{ic1}\\
&\delta\eta(\xi_B,\tau)=0,&\quad &v(\xi_B,\tau)=0.
\label{ic2}
\end{align}
\end{subequations}
%\end{eqnarray}
%
The prime at $\delta\eta$ denotes the derivative with respect to $\xi$.

Conditions (\ref{ic1}) exclude divergences at $\xi=0$ in spherical symmetry.
The two conditions in Eq.~(\ref{ic2}) ensure the DM halo radius $\xi_B$
remains unchanged, preventing changes in the chemical potential
or hydraulic pressure if the kinetic term $\sim v^2$ is added.

\begin{figure}[!t]
\centering
\captionsetup{width=0.9\textwidth}
\includegraphics[width=0.6\textwidth]{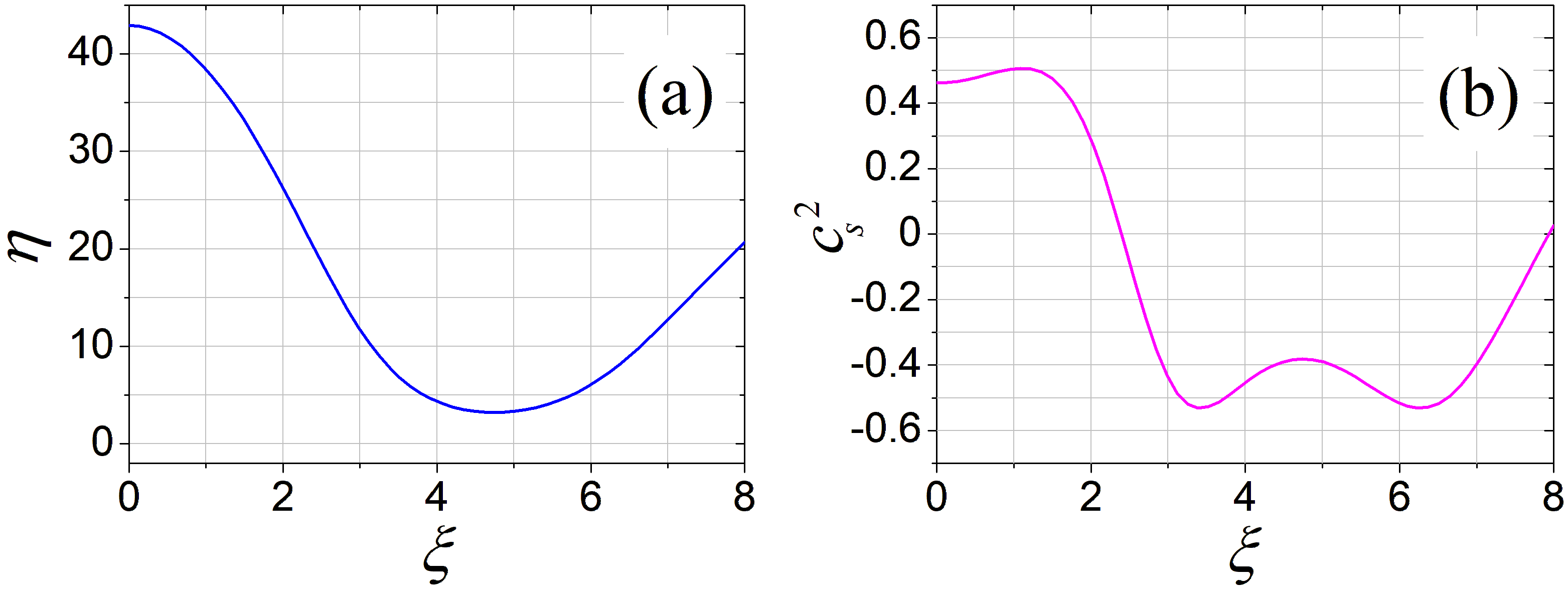} %% 8sm
%\vspace*{-1mm}
\caption{\label{ceta}Radial distribution of the background dark matter
parameters in NGC 2366 for a constant chemical potential $2u=-1.5867$.
Panel (a) illustrates the equilibrium DM number density profile $\eta(\xi)$,
smoothing out the central  cusp due to quantum pressure. Panel (b) shows
the squared speed of sound $c_s^2(\xi)$, highlighting the spatial transition
into the anomalous compressibility regime ($c_s^2<0$) in the outer
core sub-Jeans region ($\xi>2.41$).}
\end{figure}

To apply the equations to the DM in NGC~2366, we set $A=0.0045$ and
$\xi_B=6.65$, and employ the numerically obtained $\eta(\xi)$ and
$c^2_s(\xi)$ shown in Fig.~\ref{ceta}.

\subsubsection{Gas cluster expansion}

Consider the hypothetical expansion
of gas cluster-1 as it widens from the width $d_{\rm in}=0.09~\text{kpc}$
to the observed $d_{\rm fin}=0.23~\text{kpc}$ with a fixed center
$\breve{r}=1.1~\text{kpc}$, likely due to a starburst or explosion.
We associate the initial geometry with star cluster-2 in Table~\ref{tabS}
and adopt the final configuration of gas cluster-1 in Table~\ref{tabG}.
We also admit the conservation of gas mass in this process.

In general, photoionization, supernovae, and stellar winds act as stellar
feedback, heating surrounding gas and dispersing dense clouds, thereby
suppressing further star formation. Feedback from existing stars typically
reduces the efficiency of new star formation. Yet feedback can also be
positive -- shock-driven turbulence may compress cool gas, causing
the collapse of dense regions and stimulating star formation.

Carrying out the modeling in dimensionless quantities, we rewrite
the initial and final parameters of the gas cluster in units of
$r_0=1.18~\text{kpc}$:
\begin{equation}
\breve{\xi}=0.932,\ \ \ell_{\rm in}=0.076,\ \
\ell_{\rm fin}=0.195,\ \ \delta\ell_\infty=0.119.
\end{equation}
Having obtained $A_{\rm b}=0.0031$ by substituting the mass of gas cluster-1
into Eq.~(\ref{Ab}), we model the change in the baryon gravitational potential
over time as
\begin{eqnarray}
\Delta\Phi_{\rm b}(\xi,\tau)&=&\Phi_{\rm b}(\xi,\breve{\xi},\ell_{\rm fin}
-\delta\ell_\infty\rme^{-\gamma\tau})
-\Phi_{\rm b}(\xi,\breve{\xi},\ell_{\rm fin}-\delta\ell_\infty)
\nonumber\\
&\simeq&\Phi_\ell(\xi,\breve{\xi},\ell_{\rm fin})\,\delta\ell_\infty\,
(1-\rme^{-\gamma\tau}),
\\
\Delta\Phi_{\rm b}(\xi,0)&=&0,
\end{eqnarray}
where $\Phi_\ell$ appears in Eq.~(\ref{Phil}), and $\gamma$
is associated with the inverse characteristic time $1/\tau_{ch}$.

By equating $\tau_{ch}$ to the time of gravitational impact of
the expanding gas on the DM, one can roughly estimate the average velocity
of the gas:
\begin{equation}
v_{\rm gas}\sim v_0\frac{\delta\ell_\infty}{\tau_{ch}}=v_0\gamma\,\delta\ell_\infty
\simeq16.5\,\gamma~{\text{km}}/{\text{s}},
\end{equation}
where $v_0$ is taken from Eq.~(\ref{vel0}).

Velocity observations indicate that the cases $\gamma\sim5-50$
are the most likely. We set
\begin{equation}
\tau_{ch}=0.1,\qquad \gamma=10.
\end{equation}

For comparison, the gas escape velocity can be estimated from the combined
gravitational potentials of gas cluster-1 (``gc1'') and star cluster-2
(``sc2'') as infinitely thin shells. By Gauss's theorem, such
potentials coincide with the corresponding spherical potentials to give
\begin{equation}
v_{\rm esc}=\sqrt{2G\frac{M_{\rm gc1}+M_{\rm sc2}}{\breve{r}}}
\simeq13.84~{\text{km}}/{\text{s}},
\end{equation}
where $\breve{r}=\breve{r}_{\rm gc1}=\breve{r}_{\rm sc2}=1.1~\text{kpc}$
specifies the joint position of the clusters whose masses
$M_{\rm gc1}\simeq8.042\times10^6M_\odot$ and
$M_{\rm sc2}\simeq16.34\times10^6M_\odot$ are taken from Table~\ref{tabG}
and \ref{tabS}.

\begin{figure}[!t]
\centering
\includegraphics[width=0.9\textwidth]{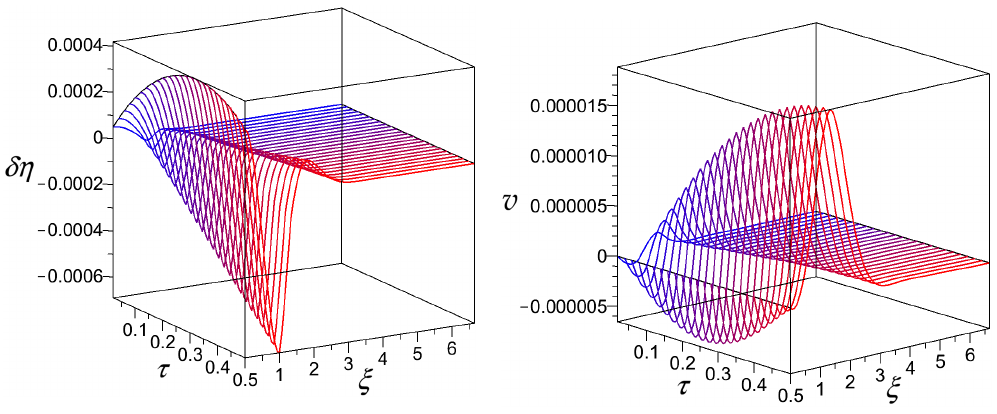}
%\vspace*{-1mm}
\caption{\label{evexp}DM perturbation density $\delta\eta$ and
 velocity $v$ during the expansion of gas cluster-1.}
\end{figure}

Since the gas expansion occurs rapidly over the time interval
$0\leq\tau\leq\tau_{ch}$, the hydrodynamics of the accompanying
DM perturbations is governed primarly by the force arising from
the gravitational potential $\delta\Phi_{\rm b}$.
Quantum fluctuations do not develop at this stage, and the velocity
profile $v$ evolves predominantly in accordance with the scaling law.
Therefore, Fig.~\ref{evexp} shows the density $\delta\eta$ and
velocity $v$ profiles already for $\tau=\tau_{ch}$. The velocity
profile describes the DM winds in opposite directions from
the cluster center, near $\breve{\xi}=0.932$, where $\delta\eta<0$.
However, this wave-like motion leads to $\delta\eta>0$ in regions
located farther from the explosion site.

Even after the baryon distribution has stabilized, DM disturbances
continue to evolve. Although the analysis for times $\tau>\tau_{ch}$
in principle requires a model adjustments, we estimate the scale of
subsequent perturbations as before, up to $\tau=5\tau_{ch}$ in
Fig.~\ref{evexp}. During this time, the velocity directed toward
the halo center reverses direction, resembling a reflection.
Simultaneously, Fig.~\ref{evexp}(left) shows an outflow of DM
particles from the halo center.

It is crucial that both the velocity $v$ (a few meters per second) and
the density deviation $\delta\eta$ (giving $\delta\eta/\eta\sim10^{-5}$)
remain small for a long time. This stability is maintained even when
disturbances arise in the region $\xi\in[2.41; 6.65]$, corresponding
to the DM unstable phase, characterized by $c^2_s<0$ in Fig.~\ref{ceta}b.
This can be attributed to the fact that quantum fluctuations develop after
the end of the gas expansion that initiated the disturbance and had a speed
exceeding the DM speed of sound $c_s$ at $\xi<2.4$. Despite the low
velocity $v$ of the DM winds, the wave-peak widths reach kiloparsec scales.
This may be due to the contribution of long-wavelength quantum fluctuations.

Note that obtaining accurate numerical solutions for high-order inhomogeneous
equations on large time scales is challenging. However, using both a centered
implicit scheme with adaptive steps and time scale variation,
$\tau^\prime=\zeta\tau$, provides reliable results and robust conclusions
for all scenarios. Furthermore, we refrained from deriving closed equations
of even higher order to avoid the influence of homogeneous solutions due to
complex kernels.

\subsubsection{Gas cluster collapse}

Let us analyze the implications of the process inverse to that considered above
-- the collapse (narrowing) of gas cluster-1, which facilitates star formation.
 
The gravitational effect of the gas mass concentration can be described using
the potential at $A_{\rm b}=0.0031$:
\begin{eqnarray}
\Delta\Phi_{\rm b}(\xi,\tau)&=&\Phi_{\rm b}(\xi,\breve{\xi},\ell_{\rm fin}
-\delta\ell_\infty\rme^{-\gamma\tau})-\Phi_{\rm b}(\xi,\breve{\xi},\ell_{\rm in})
\nonumber\\
&\simeq&-\Phi_\ell(\xi,\breve{\xi},\ell_{\rm in})\,|\delta\ell_\infty|\,(1-\rme^{-\gamma\tau}),
\end{eqnarray}
for $\delta\ell_\infty=-|\delta\ell_\infty|$, and
\begin{equation}
\breve{\xi}=0.932,\ \ \ell_{\rm fin}=0.076,\ \
\ell_{\rm in}=0.195,\ \ |\delta\ell_\infty|=0.119.
\end{equation}
This means that the average density of the gas mass increases from
$\overline{\rho}_{\rm gc}\simeq5.5\times10^{-23}~\text{kg}\,\text{m}^{-3}$
to $1.42\times10^{-22}~\text{kg}\,\text{m}^{-3}$ according to Eq.~(\ref{avrho}).
Although the final value does not yet reach the star formation threshold
in density~\cite{MK01}, we are only interested in the gravitational
perturbation created by the compaction. Stars that are accreting mass
from the surrounding gas cloud are considered protostars.

\begin{figure}[!t]
\centering
\includegraphics[width=0.9\textwidth]{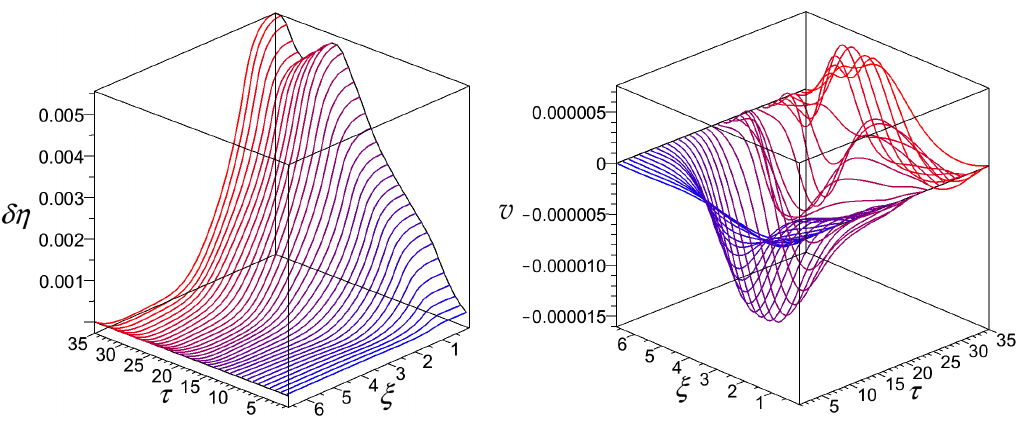}
%\vspace*{-1mm}
\caption{\label{evcoll}DM perturbation density $\delta\eta$ and
velocity $v$ during the collapse of gas cluster-1.}
\end{figure}

To estimate the timescale of gas concentration within star cluster-2
using the free-fall time of gas, we invoke the conservation of energy
expressed in dimensionless form:
\begin{equation}
\left(\frac{\rmd\xi}{\rmd\tau}\right)^2=
A_{\rm b}\left[\Phi(\xi,\breve{\xi},\ell_{\rm in})-\Phi(\xi,\breve{\xi},\ell_{\rm fin})\right],
\end{equation}
where the average initial velocity of the gas is assumed to be zero, and
the gravitational background created by star cluster-2 remains
unchanged. Then we set
\begin{equation}
\tau_{ch}=\max(\tau_{\rightarrow},\tau_{\leftarrow}),
\end{equation}
where $\tau_{\rightarrow}$ and $\tau_{\leftarrow}$ are
the moments of approaching the shell core from opposite sides:
%
%\begin{eqnarray}
\begin{subequations}
\begin{align}
&\tau_{\rightarrow}=\int_{\breve{\xi}-\ell_{\rm in}}^{\breve{\xi}}
\frac{\rmd\xi}{\sqrt{A_{\rm b}\Phi_\ell(\xi,\breve{\xi},\ell_{\rm in})\,|\delta\ell_\infty|}}
\simeq14.46,
\\
&\tau_{\leftarrow}=-\int_{\breve{\xi}+\ell_{\rm in}}^{\breve{\xi}}
\frac{\rmd\xi}{\sqrt{A_{\rm b}\Phi_\ell(\xi,\breve{\xi},\ell_{\rm in})\,|\delta\ell_\infty|}}
\simeq16.00.
\end{align}
\end{subequations}
%\end{eqnarray}

Thus, we choose
\begin{equation}
\tau_{ch}=16,\quad
\gamma=0.0625.
\end{equation}

The mean velocity of the self-gravitating gas,
\begin{equation}
v_{\rm gas}=\frac{|\delta\ell_\infty|}{\tau_{ch}}v_0\simeq1.03~\text{km}/\text{s},
\end{equation}
exceeds the sound speed of a cold molecular gas. This indicates that
gravitational forces may accidentally outweigh the supporting pressure,
leading to gas instability and its collapse according to the Jeans criteria.

When compared, the potentials $\delta\Phi_{\rm b}(\xi,\tau)$ that drive
the expansion and collapse of the same gas cluster-1 have opposite signs
and characteristic times that differ by two orders of magnitude. This
disparity implies that substantial quantum fluctuations should
develop upon reaching the moment $\tau=\tau_{ch}$ in the event of collapse.
These fluctuations, in turn, affect the formation of the perturbed density
$\delta\eta$ and velocity $v$ profiles shown in Fig.~\ref{evcoll}, obtained
by using the time variable $\tau^\prime=0.1\tau$ when integrating
the hydrodynamic equations.

By considering the time interval up to $\tau\simeq2\tau_{ch}$ in
Fig.~\ref{evcoll}, two clear trends emerge. During the slow gas collapse
the induced dark-wind velocity $v$ is roughly an order of magnitude smaller
than during the rapid gas expansion. By contrast, the perturbed density
$\delta\eta$ turns out about an order of magnitude larger in the collapse than
in the expansion, by comparing Figs.~\ref{evexp} and \ref{evcoll}.

Thus, for $\tau\leq35$ ($t\leq291~\text{Myr}$), the DM configuration remains
effectively unchanged in response to the gas cluster collapse, although
additional events should be expected on these timescales. Nevertheless,
it can be assumed that slow baryon flows induce a noticeable alteration
of the DM density.

\subsubsection{Star cluster migration}

To confirm the dominant role of slow baryonic mechanisms in DM
redistribution and to complete the picture of how stellar
displacements affect DM, we model the migration of the entire star cluster-2
towards the galactic center, driven by dynamical friction in the surrounding
gas \mbox{cluster-1.}

\begin{figure}[!t]
\centering
\includegraphics[width=0.9\textwidth]{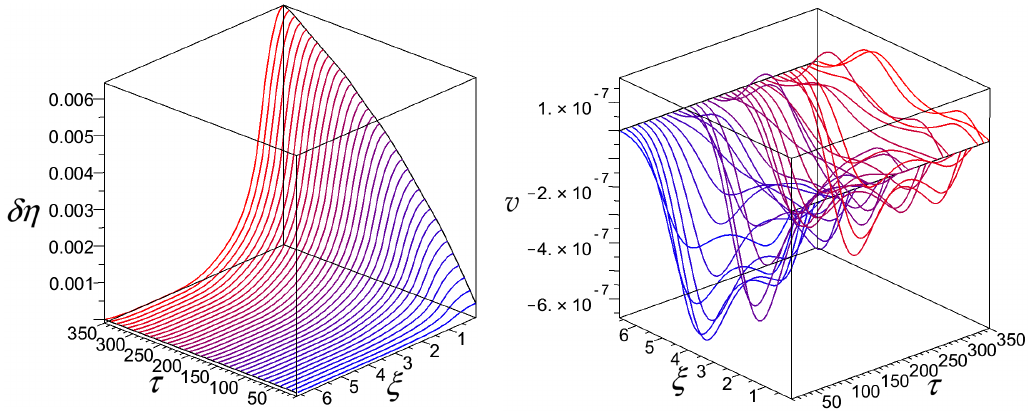}
%\vspace*{-1mm}
\caption{\label{evsm}DM perturbation density $\delta\eta$ and
velocity $v$ during the migration of star cluster-2.}
\end{figure}

Under the assumption of constant cluster mass during its migration to
the hypothetical radius $\breve{r}_{\rm fin}=0.832r_0$, the process
is characterized by the following dimensionless geometric parameters:
\begin{equation}
\breve{\xi}_{\rm in}=0.932,\ \ \ell_{\rm in}=0.076,\ \
\breve{\xi}_{\rm fin}=0.832,\ \ \delta\breve{\xi}_\infty=0.1,
\end{equation}
along with the dynamical ones
\begin{equation}
\tau_{ch}=312.5,\qquad \gamma=0.0032.
\end{equation}
To estimate a conservative upper bound $\gamma=t_0/t_{\rm df}$
for the inward migration rate in the gas cluster-1  environment
with the mean mass density
$\overline{\rho}_{\rm gc}\simeq5.5\times10^{-23}~\text{kg}\,\text{m}^{-3}$,
we use dynamical-friction scaling (\ref{tdf1}) for the mass
$M_{\rm sc}=16.34\times10^6M_\odot$ and the position
$\breve{r}=1.1~\text{kpc}$ of both clusters from Table~\ref{tabG}
and \ref{tabS}.
The rotation velocity of the stars $V_t=22~\text{km}/\text{s}$,
taken from the rotation curve of NGC~2366 in \cite{Oh08,Naz2025},
significantly exceeds the sound speed in cold gas.

Computing the gravitational parameter $A_{\rm b}=0.00622$, we modify the potential as
\begin{eqnarray}
\Delta\Phi_{\rm b}(\xi,\tau)&=&\Phi_{\rm b}(\xi,\breve{\xi}_{\rm fin}
+\delta\breve{\xi}_\infty\rme^{-\gamma\tau},\ell_{\rm in})
-\Phi_{\rm b}(\xi,\breve{\xi}_{\rm in},\ell_{\rm in})
\nonumber\\
&\simeq&-\Phi_\xi(\xi,\breve{\xi}_{\rm in},\ell_{\rm in})\,
\delta\breve{\xi}_\infty\,(1-\rme^{-\gamma\tau}).
\end{eqnarray}

The hydrodynamic evolution of DM induced by the slow migration
of star cluster-2, over timescales much longer than in previous
gas-based scenarios, is shown in Fig.~\ref{evsm} and is the result
of integrating the equations with auxiliary time variable
$\tau^\prime=0.01\tau$. This confirms
the correlation between a low rate and a large perturbation in
DM density. Furthermore, by tracking the longest process, we
observe the decay of the DM wind through a comparison of
velocity magnitudes at the beginning and end of the time
interval in Fig.~\ref{evsm}. Despite its low velocity of
the order of $1~\text{cm}/\text{s}$, the widths of its wave
peaks are enormous. We again observe an increase in wind
speed in the region of the unstable phase of DM at $\xi>2.4$.

These three scenarios reveal a profound gravitational reciprocity
between galactic components: while massive baryonic rearrangements
induce only subtle perturbations in the dominant DM substrate,
the back-reaction implies that even minor localized excitations
within the massive DM halo can disproportionately impact the lighter,
more agile baryonic matter. However, a critical physical
caveat applies: for compact baryonic sub-structures to effectively feel
this back-reaction, the DM perturbations must possess large spatial
gradients. High-gradient fluctuations in the DM density or velocity
fields, occurring on short sub-kiloparsec spatial scales, drastically
amplify local tidal forces ($F\propto\nabla\delta\Phi_{DM}$),
acting as a powerful gravitational scaffolding for gas and stars.

While the precise triggers for such high-spatial-frequency configurations
require further study, a highly plausible physical pathway is the formation
of a hydrodynamic shock wave. Such a shock naturally arises as the DM wind
transitions from the anomalous unstable phase region ($c_s^2<0$) -- where
velocity spikes are common -- into the stable phase domain characterized
by a finite, positive speed of sound. These sharp phase boundaries could
provide the exact high-gradient gravitational drive necessary to actively
compress cool gas and dictate local star-formation rates.

%% %% %% %% %% %% %% %% %% %% %% %% %% %% %% %% %% %% %% %% %% %% %% %% %% %% %% %%
\section{Multiple events and decay of excitations}\label{S4}

To evaluate the collective impact of spatially and temporally segregated subgalactic
baryonic processes on DM, we adopt the following approach. Since the relaxation of
each individual process cannot be cleanly isolated from the overarching background
of the others, we treat their effects as small perturbations, thereby suppressing
resonance phenomena and ensuring the conservation of the total DM mass.
The gravitational potentials induced by these processes are linearly superimposed
in the asymptotic limit ($t\to\infty$). Furthermore, they can be appropriately
averaged, analogous to Eq.~(\ref{1Vc}), and absorbed into a constant chemical
potential to determine the resulting quasi-stationary DM distribution.
The corresponding thermodynamic functions of DM for the NGC 2366 galaxy, evaluated
across a range of chemical potentials, have been extensively detailed in \cite{Naz2025}.

This framework inherently implies the attenuation of initial perturbations,
which rigorously justifies employing the solutions of homogeneous hydrodynamic
equations to model the subsequent evolution of the system. Given that dissipation
within an inviscid condensate can only be realized via quantum effects in
the presence of an unstable phase domain, the remainder of this Section focuses
explicitly on the underlying mechanisms of quantum relaxation.

The starting point for evolution consideration is the classical Hamiltonian
of the model, expressed in formulas (\ref{G1})-(\ref{Poi}). We recast it in
terms of the dimensionless particle number density $\eta$ and the phase
$\theta$ for the macroscopic wave function as follows:
\begin{equation}\label{Hcl}
H=\frac{1}{2}\int_0^{\xi_B}\left[\left(\partial_\xi\sqrt{\eta}\right)^2+\eta(\partial_\xi\theta)^2
+\frac{A}{2}\eta\Delta^{-1}_\xi\eta+2(1-\cos{\sqrt{\eta}})-\eta\right]\xi^2\rmd\xi.
\end{equation}
This generates the Hamiltonian equations, that is, the BEC hydrodynamic equations:
\begin{equation}
\partial_\tau\theta(\xi,\tau)=-\frac{\delta H}{\delta\eta(\xi,\tau)},\qquad
\partial_\tau\eta(\xi,\tau)=\frac{\delta H}{\delta\theta(\xi,\tau)}.
\end{equation}

Prior to formulating the quantum framework, it is essential to elucidate the physical
justification for the choice of boundary conditions. In the preceding classical
hydrodynamic treatment, the rigid-wall conditions (\ref{ic2}) served as a temporal
idealization. They simulated the robust, stabilizing external hydraulic pressure
exerted by the massive outer halo over short timescales ($\tau \sim \tau_{ch}$).
This computational constraint effectively generated standing waves, suppressing
surface modes and artificial instabilities to successfully isolate the pure forced
response of the core to baryonic redistribution. In contrast, the current Section
examines the long-term, free quantum relaxation of the entire system.
This necessitates a transition to natural Robin boundary conditions
\begin{subequations}\label{totic}
\begin{align}
&\left.\partial_\xi\delta\eta(\xi,\tau)\right|_{\xi=0}=0,&
\quad
&\left.\partial_\xi\delta\theta(\xi,\tau)\right|_{\xi=0}=0,
\label{iic1}\\
&\left.\partial_\xi\frac{\delta\eta(\xi,\tau)}{\eta(\xi)}\right|_{\xi=\xi_B}=0,&
\quad
&\left.\partial_\xi\delta\theta(\xi,\tau)\right|_{\xi=\xi_B}=0,
\label{iic2}
\end{align}
\end{subequations}
which are dictated by the smooth matching of the dense core’s density profile
with the outer halo tail.
As will be shown below, the boundary conditions (\ref{iic2}) ensure
the consistent spatial matching of the density and phase perturbation modes,
allowing the full set of constraints in (\ref{totic}) to be reduced
to the two boundary conditions specified in Eq.~(\ref{fiic}) for the auxiliary
eigenvalue problem.

Substituting $\eta(\xi,\tau)\to\eta(\xi)+\delta\eta(\xi,\tau)$ and
$\theta(\xi,\tau)\to-u\tau+\theta(\xi,\tau)$, we extract from $H$
the contribution quadratic in the perturbations $\delta\eta$ and
$\theta$, that is, the second variation:
\begin{equation}
H^{(2)}=\frac{1}{2}\int_0^{\xi_B}\left[-\frac{\delta\eta}{4}\widetilde{\Delta}_\xi\frac{\delta\eta}{\eta}
-\eta\theta\widetilde{\Delta}_\xi\theta+\frac{A}{2}\delta\eta\Delta^{-1}_\xi\delta\eta
+\frac{c^2_s}{2\eta}(\delta\eta)^2\right]\xi^2\rmd\xi.
\end{equation}

Due to the equations of motion (\ref{HEqs}) without a baryon source,
the quadratic Hamiltonian is reduced to
\begin{equation}
H^{(2)}=\frac{1}{2}\int_0^{\xi_B}(\theta\,\partial_\tau\delta\eta-\delta\eta\,\partial_\tau\theta)
\,\xi^2\rmd\xi.
\end{equation}

To describe the evolution of flows when velocity and phase are
related as $v=\partial_\xi\theta$, we turn to the homogeneous equation:
\begin{equation}\label{wt}
2\partial^2_\tau\theta(\xi,\tau)-c_s^2(\xi)\widetilde{\Delta}_\xi\theta(\xi,\tau)
+\frac{\widetilde{\Delta}_\xi^2}{2}\theta(\xi,\tau)
-A\Delta_\xi^{-1}\left[\eta(\xi)\widetilde{\Delta}_\xi\theta(\xi,\tau)\right]=0,
\end{equation}
where the hydrodynamic Laplace operator $\widetilde{\Delta}_\xi$
from Eq.~(\ref{hDel}) is used.

For the steady state with the particle density $\eta(\xi)$ and
square of sound speed $c^2_s(\xi)$, $\theta=-u\tau$ serves as a solution
to Eq.~(\ref{wt}) at constant chemical potential $u$ in dimensionless units.

To illustrate the expected behavior, consider a uniform, spherically 
symmetric system with a constant density $\eta=\text{const}$ and speed 
of sound $c_s$. In this case, the modified Laplace operator 
$\widetilde{\Delta}_\xi$ reduces to the ordinary radial Laplacian, 
$\widetilde{\Delta}_\xi=\Delta_\xi$. The solution to Eq.~(\ref{wt}) 
is then obtained by separating variables for each $n$-th mode, 
$\exp{(\pm\rmi\omega_n\tau)}\,f_n(\xi)$, within the finite interval 
$\xi \in [0; \xi_B]$. This yields the orthonormal spherical waves:
\begin{equation}\label{funn}
f_0(\xi)=\sqrt{\frac{3}{\xi_B^3}},\qquad
f_{n>0}(\xi)=\sqrt{\frac{2}{\xi_B^3}}\frac{j_0(k_n\xi)}{j_0(k_n\xi_B)},\qquad
k_n=\frac{a_n}{\xi_B},
\end{equation}
where $a_n$ is the $n$-th zero of function $j_0^\prime(a_n)=-j_1(a_n)=0$.

Given $\Delta_\xi f_n(\xi)=-k_n^2f_n(\xi)$, the corresponding dispersion relation
reproduces the Bogolyubov quasiparticle spectrum in dimensionless variables:
\begin{equation}
\omega_n=\pm\sqrt{\frac{k_n^4}{4}+\frac{1}{2}(c^2_sk_n^2-A\eta)}.
\end{equation}
The excitation frequency $\omega_n$ remains real for wave numbers exceeding 
a certain threshold, $k_n>k_{\rm th}$~\cite{H2019,Ch2020}. Conversely, 
for $k_n<k_{\rm th}$, the spectrum becomes purely imaginary, signaling 
the onset of either unstable or metastable states. Characterizing these states 
is crucial for our subsequent analysis; as shown below, what initially 
appears to be a physical limitation provides a self-consistent mechanism 
to resolve the underlying relaxation problem.

Thus, we need to find the eigenvectors and eigenvalues of the operator
\begin{equation}\label{LOp}
\widehat{L}=\frac{1}{2}\widetilde{\Delta}_\xi^2
-c_s^2(\xi)\widetilde{\Delta}_\xi
-A\Delta_\xi^{-1}\eta(\xi)\widetilde{\Delta}_\xi.
\end{equation}

It is advisable to first study the spectrum of the hydrodynamic Laplacian
$\widetilde{\Delta}_\xi$:
\begin{equation}\label{fii}
\widetilde{\Delta}_\xi\frac{\varphi_n(\xi)}{\sqrt{\eta(\xi)}}=
-\kappa_n^2\frac{\varphi_n(\xi)}{\sqrt{\eta(\xi)}},
\end{equation}
by imposing boundaries according to (\ref{iic1})-(\ref{iic2}):
\begin{equation}\label{fiic}
\varphi^\prime_n(0)=0,\qquad
\left.\partial_\xi\frac{\varphi_n(\xi)}{\sqrt{\eta(\xi)}}\right|_{\xi=\xi_B}=0.
\end{equation}

\begin{figure}[!t]
\centering
\captionsetup{width=0.9\textwidth}
\includegraphics[width=0.7\textwidth]{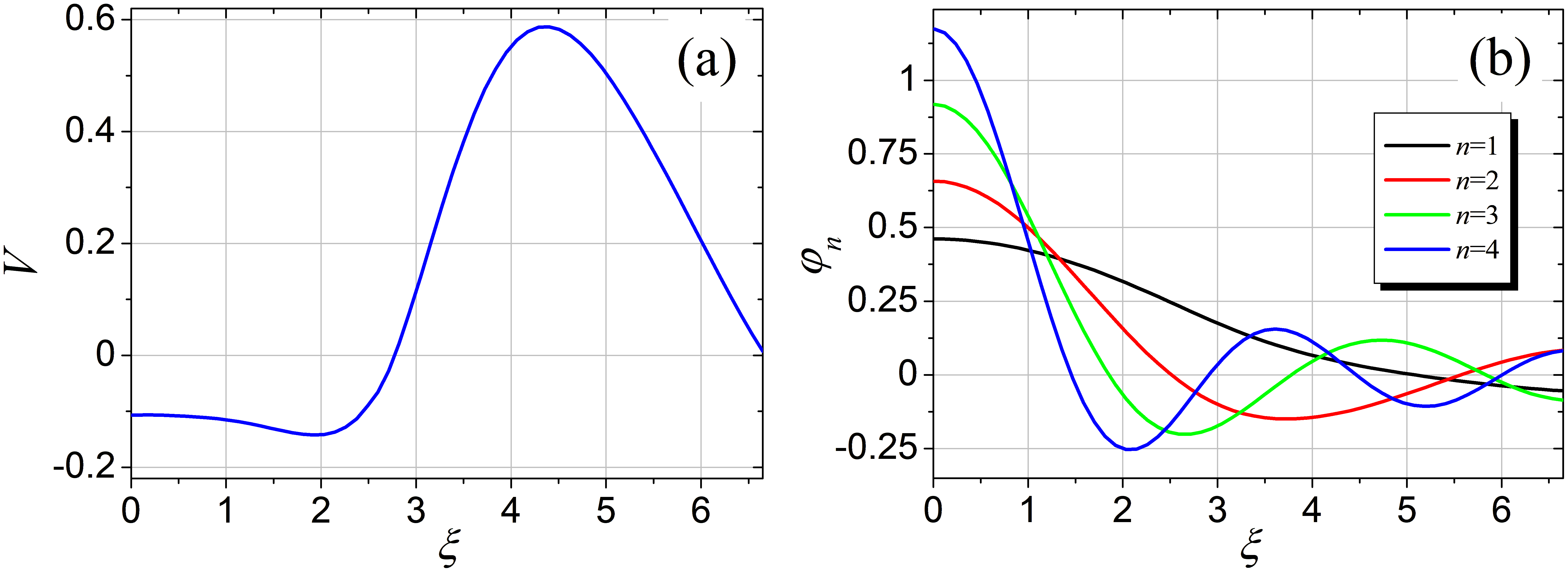}
%\vspace*{-1mm}
\caption{\label{lap}Quantum-mechanical functions under Robin boundary
conditions at $\xi_B = 6.65$. Panel (a) displays the effective
self-generated quantum pressure potential $V(\xi)$, revealing a finite
potential well for $\xi<2.75$ bounded by a semi-permeable barrier that
regulates metastable states. Panel (b) depicts the first four 
orthonormal radial wave functions $\varphi_n(\xi)$ ($n=1,2,3,4$), where
$n$ specifies the number of nodes within the spatial domain.}
\end{figure}

This equation is equal to the auxiliary problem:
\begin{eqnarray}
&&
\left[-\Delta_\xi+V(\xi)\right]\varphi_n(\xi)=\kappa_n^2\varphi_n(\xi),
\label{hlap}\\
&&
V(\xi)\equiv\frac{1}{\sqrt{\eta(\xi)}}\Delta_\xi\sqrt{\eta(\xi)},
\end{eqnarray}
where the potential $V$ coincides with the quantum pressure term
and is shown in Fig.~\ref{lap}a.

The potential $V$ possesses a well of finite depth within $\xi\in[0; 2.75]$,
bounded on the right by a barrier of finite height.
Such a configuration supports metastable states, which may decay through
quantum tunneling across the barrier. Metastability is also confirmed
by other potentials in this system, shown in Fig.~7 in \cite{Naz2025}.

The Eq.~(\ref{hlap}) is equivalent to the stationary Schr\"odinger
equation with positive energy. In this regime, a particle subject
to the finite potential $V$, confined within the spatial interval
$\xi\in[0;\xi_B]$, should experience its influence less and less as
the energy $\kappa_n^2$ increases.

Solving this, we intend to expand the functions $\{\varphi_n\}$ in some
known basis $\{\Psi_n\}$:
\begin{equation}
\varphi_m(\xi)=\sum\limits_{n}A_{m,n}\Psi_n(\xi).
\end{equation}

\begin{table}[t]
\centering
\captionsetup{width=12cm}
\caption{The values of $k_n=b_n/\xi_B$ and the eigenvalues of operators.}
\begin{tabular}{|c|c|c|c|}
\hline
 $n$ & $k_n$  &  $\kappa_n$ & $\omega_n$ \\
\hline
  1 & 0.6032685 & 0.6336482 & -i0.244451 \\
  2 & 1.1178011 & 1.2061777 & 0.6076448 \\
  3 & 1.6093111 & 1.6575698 & 1.2768967 \\
  4 & 2.0927479 & 2.1317883 & 2.1915262 \\
  5 & 2.5724626 & 2.6037764 & 3.3145550 \\
  6 & 3.0501607 & 3.0764528 & 4.6601705 \\
  7 & 3.5266448 & 3.5493894 & 6.2289728 \\
  8 & 4.0023420 & 4.0222928 & 8.0206305 \\
  9 & 4.4775000 & 4.4953537 & 10.036171 \\
 10 & 4.9522726 & 4.9683823 & 12.275179 \\
 11 & 5.4267602 & 5.4414535 & 14.737926 \\
 12 & 5.9010309 & 5.9145466 & 17.424538 \\
\hline
\end{tabular}
\label{tabE}
\end{table}

We define the radial orthonormal functions $\{\Psi_n\}$ as
\begin{equation}
\Psi_{n}(\xi)=\sqrt{\frac{2}{\xi_B^3}}\,\frac{b_{n}}
{j_0(b_{n})\sqrt{b^2_{n}+h(h-1)}}\,
j_0\left(b_{n}\frac{\xi}{\xi_B}\right),
\end{equation}
where the discrete set of positive numbers $\{b_n \mid 0 < b_1 < b_2 < \dots\}$ 
is parametrized by $h$ and determined as the successive roots of the 
Robin boundary condition:
\begin{equation}
b_{n} j^\prime_0(b_{n})+h j_0(b_{n})=0.
\end{equation}

The value of $h$ is directly derived from Eq.~(\ref{ic2}) to give us
\begin{equation}
h\equiv-\left.\xi\frac{\partial_\xi\eta(\xi)}{2\eta(\xi)}\right|_{\xi=\xi_B}
\simeq-2.42.
\end{equation}

By numerically calculating the eigenvectors and values of the symmetric matrix
\begin{equation}
D_{n,m}=\int_0^{\xi_B}\Psi_n(\xi)\left[\Delta_\xi-V(\xi)\right]\Psi_m(\xi)\,\xi^2\rmd\xi
=-\left(\frac{b_n}{\xi_B}\right)^2\,\delta_{n,m}-V_{n,m},
\end{equation}
we obtain the orthonormal wave functions $\varphi_n$, see Fig.~\ref{lap}b,
and the wave numbers $\kappa_n$ in the medium, see Table~\ref{tabE}, the
total number of which we limit to $n_{\max}=32$ in practice.

By construction, orthogonality of the set $\{\varphi_n\}$ is exact:
\begin{equation}\label{orff}
\delta_{n,m}=\int_0^{\xi_B}\varphi_n(\xi) \varphi_m(\xi)\,\xi^2\rmd\xi.
\end{equation}

The matrix element of the pseudo-Hermitian operator $\widehat{L}$ from
Eq.~(\ref{LOp}) in the basis $\{\varphi_n\}$ is
\begin{eqnarray}
(\widehat{L})_{m,n}&\equiv&\int_0^{\xi_B}\rmd\xi\,\xi^2\eta
\left(\frac{\varphi_m}{\sqrt{\eta}}\right)
\widehat{L}\left(\frac{\varphi_n}{\sqrt{\eta}}\right)
\nonumber\\
&=&\frac{\kappa^4_n}{2}\delta_{m,n}+\kappa_n^2 
\int_0^{\xi_B}\left[\varphi_m c^2_s\varphi_n
+A(\varphi_m\sqrt{\eta})\Delta^{-1}_\xi(\sqrt{\eta}\varphi_n)\right]\xi^2\rmd\xi.
\end{eqnarray}
The excitation spectrum $\{\omega_n\}$ is subsequently determined
by solving the characteristic equation
$\det\|(\widehat{L})_{m,n}-2\omega^2\delta_{n,m}\|=0$.

Using the Hermitian matrices 
$\widehat{Q}=\left\|\kappa_n^{-1}\delta_{n,m}\right\|$
and $\widehat{Q}=\left\|\kappa_n\delta_{n,m}\right\|$,
the pseudo-Hermitian matrix $\widehat{L}$ is transformed
into a Hermitian one as
\begin{equation}
\widehat{L}_Q=\widehat{Q}^{-1}\widehat{L}\widehat{Q}.
\end{equation}
This saves the spectrum in Table~\ref{tabE}, but the eigenvectors
${\vec u}_n=(B_{m,n})$ of the symmetric matrix $\widehat{L}_Q$
provide the orthogonality of the normalized wave functions:
\begin{equation}
\psi_m(\xi)=\sum\limits_{n} B_{m,n} \varphi_n(\xi).
\end{equation}

The following properties are crucial:
\begin{equation}\label{BB}
\sum\limits_{n} B_{n,m}B_{n,k}=\delta_{m,k},
\qquad
\sum\limits_{n} B_{m,n}B_{k,n}=\delta_{m,k}.
\end{equation}

The set of non-orthogonal eigenfunctions $\widetilde{\psi}_n$ of
the operator $\widehat{L}$ is then calculated as
\begin{equation}\label{wtow}
\widetilde{\psi}_m(\xi)=\sum\limits_{n} \frac{B_{m,n}}{\kappa_n} \varphi_n(\xi).
\end{equation}

The central issue at this stage is to express the spectrum
in the form:
\begin{equation}\label{dpr}
\omega^2(\kappa)=\frac{\kappa^4}{4}+\alpha(\kappa)\frac{\kappa^2}{2},\qquad
\omega_n=\omega(\kappa_n),
\end{equation}
where $\alpha(\kappa)$ takes into account the contribution
of the non-uniform speed of sound and gravity for the wave
number $\kappa$ in the medium.

\begin{figure}[t]
\centering
\captionsetup{width=0.9\textwidth}
\includegraphics[width=0.7\textwidth]{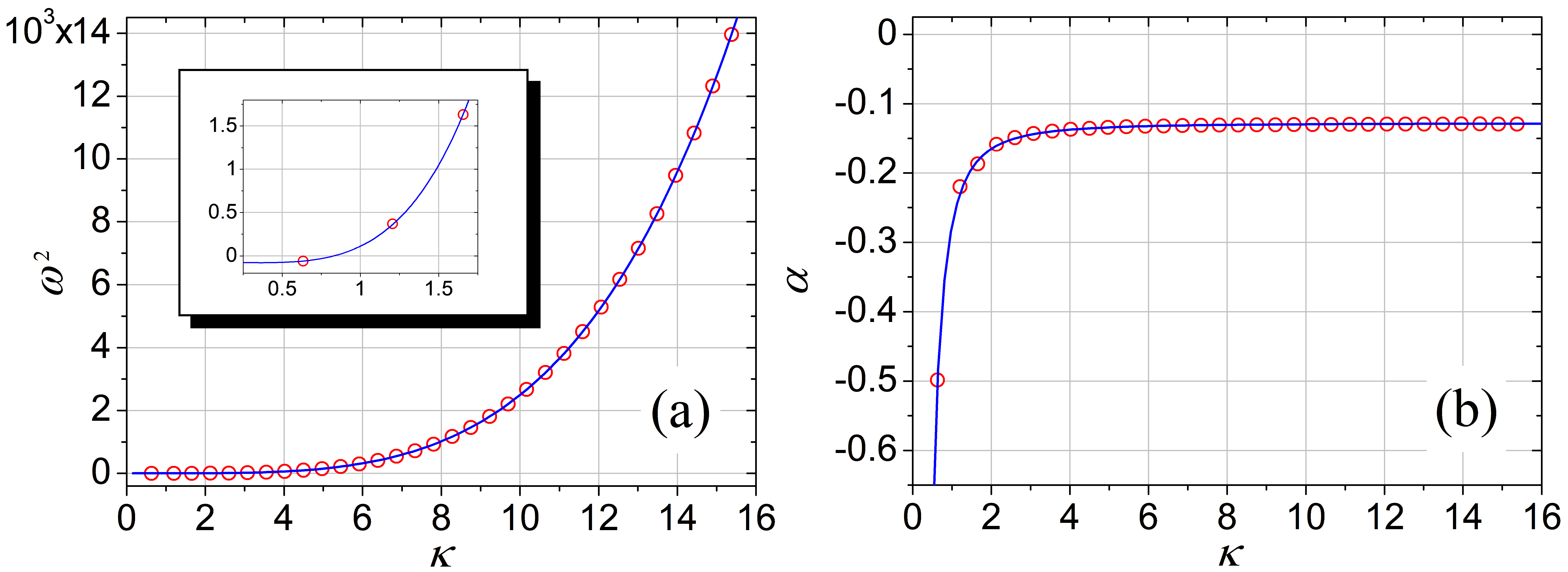}
\caption{\label{lmbd}Excitation spectrum of the
inhomogeneous self-gravitating BEC DM halo. Panel (a) shows the
squared collective frequencies $\omega_n^2$ versus the wave number
$\kappa_n$, where circles denote discrete matrix-diagonalization
eigenvalues and the blue line marks the analytical fit. The inset
highlights the stable-unstable crossover at $\kappa \simeq 0.85$. 
Panel (b) presents the non-uniform speed of sound and gravity
correction function $\alpha(\kappa)$ computed via relation (\ref{dpr}) 
alongside its optimized multi-parameter fit $\alpha_{\text{fit}}(\kappa)$
(\ref{afit}).}
\end{figure}

The unstable DM region, characterized by $c^2_s(\xi)<0$, together with
the self-gravity of DM, leads to negative values of
$\alpha\in[-0.498; -0.1294]$ in the given range of $\kappa$.
As $\kappa$ increases, $\alpha(\kappa)$
rapidly approaches its maximal value $\alpha_{\max}\simeq-0.1294$,
as shown in Fig.~\ref{lmbd}b.
Given that gravity makes a contribution $\sim\kappa^{-2}$,
the best fit to $\alpha(\kappa)$ is
\begin{equation}\label{afit}
\alpha_{\rm fit}(\kappa)=-0.1282-\frac{0.1477}{\kappa^2}.
\end{equation}

Examining the dependence $\omega^2(\kappa)$ in Fig.~\ref{lmbd}a,
we see that the modes are stable at $\kappa\geq0.85$, when
$\omega^2(\kappa)>0$.
Within the strict framework of spherical
symmetry maintained throughout this work, the lowest discrete
long-wavelength mode $\omega _{1}$ (at orbital number $l=0$)
emerges as the single purely imaginary eigenvalue in Table~\ref{tabE}.
This lone unstable mode captures the essential physics of Jeans-like
destabilization inside the medium. While non-spherical perturbations could
potentially trigger higher multipole instabilities, they lie outside
our current geometric ansatz and are effectively frozen. This allows
us to use a single-mode imaginary spectrum as a base value for
estimating the false vacuum lifetime $t_\text{vd}$ below.

Combining, a solution to Eq.~(\ref{wt}) can be written in the form~\cite{LP80}:
\begin{equation}\label{thx}
\theta(\xi,\tau)=\sum\limits_{n,m}B_{n,m}
\left(\frac{\omega_n}{2\eta(\xi)\kappa_m^2}\right)^{1/2}
\left(\beta_n\rme^{-\rmi\omega_n\tau}+\overline{\beta}_n\rme^{\rmi\omega_n\tau}\right)\varphi_m(\xi),
\end{equation}
where we substituted Eq.~(\ref{wtow}); $\beta_n$ and $\overline{\beta}_n$
are complex conjugate amplitudes, which are defined by initial conditions.

Due to the continuity relation
$\partial_{\tau}\delta\eta=-\eta\,\widetilde{\Delta}_\xi\theta$
and Eq.~(\ref{fii}), we derive the density perturbation:
\begin{equation}\label{etx}
\delta\eta(\xi,\tau)=\rmi\sum\limits_{n,m}B_{n,m}
\left(\frac{\eta(\xi)\kappa_m^2}{2\omega_n}\right)^{1/2}
\left(\beta_n\rme^{-\rmi\omega_n\tau}-\overline{\beta}_n\rme^{\rmi\omega_n\tau}\right)\varphi_m(\xi).
\end{equation}

Furthermore, we promote the amplitudes to bosonic creation and
annihilation operators, $\overline{\beta}_n\to\hat{\beta}^\dag_n$
and $\beta_n\to\hat{\beta}_n$, which satisfy the canonical
commutation relation:
\begin{equation}
[\hat{\beta}_n,\hat{\beta}^\dag_m]\equiv
\hat{\beta}_n\hat{\beta}^\dag_m-\hat{\beta}^\dag_m\hat{\beta}_n
=\delta_{n,m}.
\end{equation}

Using Eq.~(\ref{BB}), the commutator of quantized fields at the same
$\tau$ is reduced to
\begin{equation}
\left[\widehat{\theta}(\xi_1,\tau),\widehat{\delta\eta}(\xi_2,\tau)\right]
=-\rmi\sum\limits_{n}\varphi_n(\xi_1)\varphi_n(\xi_2)
=-\frac{\rmi}{\xi_1\xi_2}\delta(\xi_1-\xi_2),
\end{equation}
where the last equality is valid for the complete set $\{\varphi_n\}$,
when $n_{\max}\to\infty$.

The form of fields (\ref{thx}) and (\ref{etx}), as well as their versions
with quantized amplitudes, could also be obtained using the Bogolyubov
transform, taking into account the presence of unstable modes as in
\cite{Iz25,Mine,Kob08}.

Upon substituting the quantized perturbations $\widehat{\delta\eta}$
and $\widehat{\theta}$, and invoking the orthogonality of
$\{\varphi_n\}$ and $\{B_{n,m}\}$ as given in Eqs.~(\ref{orff}) and (\ref{BB}),
we formally obtain the canonical quantum Hamiltonian:
\begin{equation}\label{hH2}
\widehat{H}^{(2)}=\frac{1}{2}\sum\limits_n\omega_n
\left(\hat{\beta}_n\hat{\beta}^\dag_n+\hat{\beta}^\dag_n\hat{\beta}_n\right)
=\sum\limits_n\omega_n\left(\hat{\beta}^\dag_n\hat{\beta}_n+\frac{1}{2}\right).
\end{equation}

Despite its formal resemblance to the Bogolyubov Hamiltonian for excitations,
their spectrum $\omega_n=\omega(\kappa_n)$ exhibits distinctive features.
In particular, it does not generate phonon modes at small $\kappa$, which
allows it to be interpreted rather as a single-particle spectrum in the medium.
In essence, it is governed by the dispersion relation:
\begin{equation}
\omega(\kappa)\simeq\frac{1}{2}\left(\kappa^2+\alpha(\kappa)\right),\qquad
\kappa\gg1.
\end{equation}
At small $\kappa$, the modes cannot even be associated with quasiparticles
when tunneling through the $V$-barrier, as shown in Fig.~\ref{lap}.

Moreover, the lowest mode contributes purely imaginary term into $\widehat{H}^{(2)}$:
\begin{eqnarray}
&&\widehat{H}^{(2)}=\widehat{H}_0-\rmi\widehat{\Gamma}_0,
\\
&&\widehat{H}_0=\sum\limits_{n\geq2}\omega_n\left(\hat{\beta}^\dag_n\hat{\beta}_n+\frac{1}{2}\right),
\quad
\widehat{\Gamma}_0=|\omega_1|
\left(\hat{\beta}^\dag_1\hat{\beta}_1+\frac{1}{2}\right),
\quad
[\widehat{H}_0,\widehat{\Gamma}_0]=0.
\end{eqnarray}
The minus sign preceding $\widehat{\Gamma}_0$ is introduced to ensure
that Hamiltonian (\ref{hH2}) is recovered under the following identification
$|\omega_1|=\rmi\omega_1$, where $\rmi=\sqrt{-1}$ as usual. This is also
dictated by the properties of the scattering matrix.

While $\widehat{H}_0$ acts naturally on the Fock space spanned by the occupation
number basis, the state space for $\widehat{\Gamma}_0$ is not a simple Fock
one~\cite{Mine,Kob08}. The specific eigenstates of $\widehat{\Gamma}_0$, defined
in \cite{Kob08}, allow us to retain the Hamiltonian structure and notations
in the presence of purely imaginary mode. We also stress the projection
method developed in \cite{Iz25} for the BEC with unstable modes.

Thus, the steady BEC background $\eta(\xi)$ for $\xi\in[0; \xi_B]$, where $\xi_B$
is smaller than the Jeans radius, gives rise to the discrete long-wavelength
mode with imaginary cyclic frequency. This spectrum reflects a metastable configuration
reminiscent of a false vacuum~\cite{Col77,Sh96,YSL}: locally stable but unstable
to large-scale perturbations. Although the unstable mode can be eliminated
from the density and conjugate phase perturbations, the quantum energy expression
should retain a residual vacuum contribution.

By setting $\hat{\beta}^\dag_1=0$ and $\hat{\beta}_1=0$,
the vacuum contribution enters as a constant
\begin{equation}
\gamma=\frac{1}{2}\,|\omega_1|\simeq0.1222,
\end{equation}
which represents the vacuum imprint of this sector.

Due to commutativity of the components, the resulting evolution operator
factorizes as
\begin{equation}
\rme^{-\rmi\widehat{H}_0\tau}\rme^{-\gamma\tau} \quad\text{for}\quad
\tau\geq0,
\end{equation}
where the first factor governs the unitary dynamics of the stable
quasiparticle modes, while the second factor encodes a universal
exponential damping. This structure reflects the intrinsic instability
of the quasiparticle vacuum: even excitations in the stable
sector decay in time due to the background instability.
However, taking into account the specifics of the system,
the dependence of the modes on time remains the same as in
Eqs.~(\ref{thx}), (\ref{etx}).

Notably, these unstable modes break time-reversal invariance ($\tau\to-\tau$). 
Their behavior in non-inertial reference frames, specifically regarding 
the Unruh effect, has been systematically investigated in~\cite{FP25}.
This connection may be particularly significant given the role of gravity
in their formation.

The relaxation rate due to the decay of metastable state 
is estimated as
\begin{equation}
t_{\rm vd}=\frac{t_0}{\gamma}\simeq6.8\times10^7~\text{yr}.
\end{equation}
Thus, hydrodynamic perturbations should weaken over
time $t\geq t_{\rm vd}$, bringing the system to a steady state described
by equilibrium $\eta(\xi)$ and $\theta=-u\tau$.
Furthermore, $t_{\rm vd}$ is much smaller than the ``infinite-time'' limit,
which is on the order of $10^9$ years.

To rigorously demonstrate that the configuration with a purely imaginary frequency
$\omega_1=-\rmi|\omega_1|$ constitutes an unstable saddle point rather than
a mathematical artifact of the linear approximation, we must extend our analysis
by including the third- and fourth-order Hamiltonian corrections, $H^{(3)}$ and $H^{(4)}$.

At the classical level, focusing strictly on the lowest spherically symmetric radial
mode ($l=0, n=1$) and frozen higher multipoles ($l \neq 0$),
we turn to higher-order variations of the following effective forms:
\begin{eqnarray}
&&\hspace{-5mm}
H^{(3)}=\frac{1}{2}\int_0^{\xi_B}
\left[\frac{(\delta\eta)^2}{8\eta}\widetilde{\Delta}\frac{\delta\eta}{\eta}
-\delta\eta\,\theta \widetilde{\Delta}\theta
+\frac{\eta}{2} \theta^2\widetilde{\Delta}\frac{\delta\eta}{\eta}
+j_2\left(\sqrt{\eta}\right)\frac{(\delta\eta)^3}{24\,\eta}
\right]\xi^2\rmd\xi,
\\
&&\hspace{-5mm}
H^{(4)}=\frac{1}{2}\int_0^{\xi_B}
\left[
-\frac{(\delta\eta)^3}{16\,\eta^2}\widetilde{\Delta}\frac{\delta\eta}{\eta}
-j_3\left(\sqrt{\eta}\right)\frac{(\delta\eta)^4}{192\,\eta^{3/2}}
\right]\xi^2\rmd\xi.
\end{eqnarray}

To construct the real energy functional analogous to the Coleman potential~\cite{Col77},
we map the quantized field amplitudes $a$ and ${\bar a}$, such that
${\hat\beta}_{1}|a\rangle=a|a\rangle$, to the new real canonical variables $(Q, P)$:
\begin{equation}
Q=\rmi\rme^{\rmi\pi/4}\,\frac{a-{\bar a}}{\sqrt{2}},\qquad
P=\rme^{-\rmi\pi/4}\,\frac{a+{\bar a}}{\sqrt{2}},
\end{equation}
which satisfy the standard Poisson bracket $\{Q, P\}=1$ and formally yield 
\begin{equation}\label{qH}
H^{(2)}=\frac{|\omega_1|}{2}\left(P^2-Q^2\right)\quad\text{for}\quad
|a|^2=\frac{\rmi}{2}\left(P^2-Q^2\right).
\end{equation}

Treating the variables $Q$ and $P$ as \textit{real} is a physically
justified step; it preserves the number of degrees of freedom, but prevents
a return to the initial amplitudes. This reflects the lack of an isomorphism
between the algebras $so(2)$ and $so(1,1)$, as well as the transition to
a hyperbolic phase space, in which the origin becomes an unstable saddle point.

Considering only the contribution of the unstable mode to the perturbations
(\ref{thx}), (\ref{etx}),
\begin{equation}
\theta(\xi,\tau)=\sqrt{\frac{|\omega_1|}{\eta(\xi)}}\,P(\tau)\,F^{(-1)}(\xi),\quad
\delta\eta(\xi,\tau)=\sqrt{\frac{\eta(\xi)}{|\omega_1|}}\,Q(\tau)\,F^{(1)}(\xi),\quad
F^{(s)}(\xi)\equiv\sum\limits_{n}B_{1,n}\,\kappa_{n}^s\,\varphi_{n}(\xi),
\end{equation}
the resulting full non-linear Hamiltonian, $H^{(2)}+H^{(3)}+H^{(4)}$,
for the single-mode subsystem reads:
\begin{equation}\label{Hv}
H(Q,P)=\frac{|\omega_1|}{2}\left(P^2-Q^2\right)
+\Lambda_3\,Q^3+\mu_3\,Q\,P^2+\Lambda_4\,Q^4,
\end{equation}
where the self-interaction couplings are defined by the radial overlap integrals:
\begin{eqnarray}
&&
\Lambda_3=-\frac{1}{16\,|\omega_1|^{3/2}}
\int_0^{\xi_B}\left[F^{(3)}-\frac{\eta}{3}\,j_2\left(\sqrt{\eta}\right)\,F^{(1)}\right]
\,\left(F^{(1)}\right)^2\,\frac{\xi^2\,\rmd\xi}{\sqrt{\eta}},
\label{I1l}\\
&&
\mu_3=\frac{\sqrt{|\omega_1|}}{4}\,
\int_0^{\xi_B}\left[2F^{(1)}\,F^{(1)}-F^{(3)}\,F^{(-1)}\right]
\,F^{(-1)}\,\frac{\xi^2\,\rmd\xi}{\sqrt{\eta}},
\label{I2l}\\
&&
\Lambda_4=\frac{1}{32\,|\omega_1|^{2}}\int_0^{\xi_B}
\left[F^{(3)}-\frac{\eta^{3/2}}{12}\,j_3\left(\sqrt{\eta}\right)\,F^{(1)}\right]
\,\left(F^{(1)}\right)^3\,\frac{\xi^2\,\rmd\xi}{\eta}.
\label{I3l}
\end{eqnarray}
These coefficients take values
$\Lambda_3=-1.530\times10^{-3}$, $\mu_3=-4.104\times10^{-3}$, and $\Lambda_4=1.266\times10^{-5}$.

\begin{figure}[!t]
\centering
\captionsetup{width=0.9\textwidth}
\includegraphics[width=0.7\textwidth]{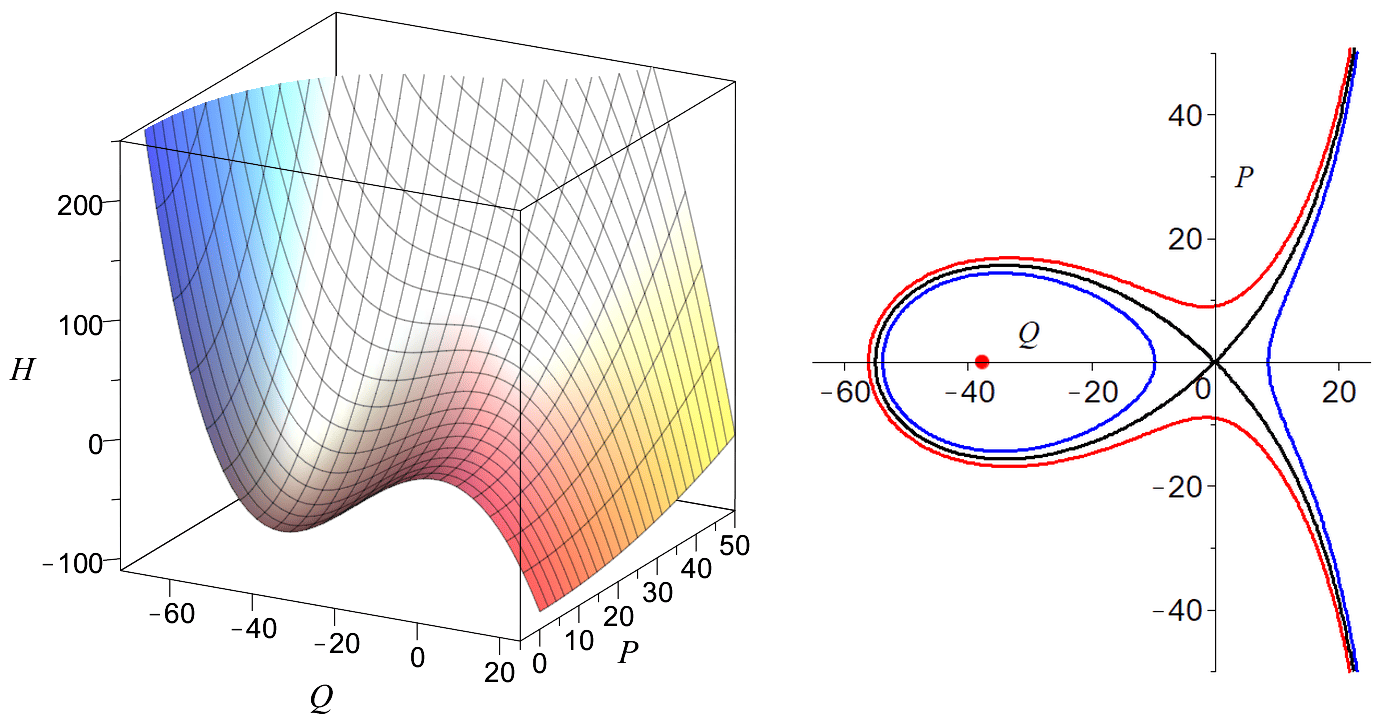} %% 8sm
%\vspace*{-1mm}
\caption{\label{Hp}Left panel:  Extrema of the Hamiltonian $H(Q,P)$ from (\ref{Hv}).
Right panel: Phase portrait of the trajectories
$H=\text{const}$. Red curve: $H>0$. Black curve: $H=0$. Blue curve: $H<0$.
Red dot indicates the local minimum at $(Q=Q_*,P=0)$.}
\end{figure}

The form of the effective potential $V_{\rm eff}(Q)=H(Q, 0)$ clearly reveals
the nature of the false vacuum transition (see Fig.~\ref{Hp}). The origin $(Q=0, P=0)$
represents a true unstable saddle point since the Hessian matrix exhibits mixed signs:
$\partial^2 H/\partial P^2=|\omega_1|>0$ and $\partial^2 H/\partial Q^2=-|\omega_1|<0$.
By finding the extrema from the conditions $\partial H/\partial Q=0$, we conclude
that the system rolls down from the false vacuum along the tachyonic direction into
a stable, true vacuum configuration located at $Q_*\simeq-37.62$,
where the local parabolic well robustly stabilizes the phase-space trajectories into
closed elliptic orbits.

In an infinite or continuous medium, the cubic variation $H^{(3)}$
typically describes the standard Beliaev damping~\cite{LP80,Bel2,MVS,CB09,P15},
representing the irreversible decay of a quasiparticle into a continuum of
lower-energy modes. However, in our finite-volume system characterized by
a discrete and non-equidistant energy spectrum, exact simultaneous conservation
of energy and momentum for a single-particle decay into an open continuum cannot
be satisfied. Consequently, the third-order Hamiltonian term mediates a nonlinear, 
three-mode parametric energy exchange, which is analogous to wave-mixing processes
in nonlinear optics.

The matrix elements
\begin{equation}
T_{m,n,k}\equiv\int_0^{\xi_B}\varphi_m \varphi_n \varphi_k\,\frac{\xi^2\rmd\xi}{\sqrt{\eta}},
\qquad
S_{m,n,k}\equiv\int_0^{\xi_B}\sqrt{\eta}\,j_2(\sqrt{\eta})\,\varphi_m
\varphi_n\varphi_k\,\xi^2\rmd\xi,
\end{equation}
so that
\begin{equation}
\int_0^{\xi_B}\sqrt{\eta}\varphi_m\left(\partial_\xi\frac{\varphi_n}{\sqrt{\eta}}\right)
\left(\partial_\xi\frac{\varphi_k}{\sqrt{\eta}}\right)\,\xi^2\rmd\xi
=\frac{\kappa_k^2+\kappa_n^2-\kappa_m^2}{2}\,T_{m,n,k},
\end{equation}
define the effective couplings between these discrete localized states, 
facilitating a cascade-like energy transfer from high-frequency excitations 
down to long-wavelength modes, thereby providing a purely quantum 
mechanism for the relaxation of spatial inhomogeneities over time.
However, one might expect a minor effect arising from such
an interaction at the next order of perturbation theory.
A detailed study of this issue will be carried out elsewhere.

%% %% %% %% %% %% %% %% %% %% %% %% %% %% %% %% %% %% %% %% %% %% %% %% %% %% %% %%
\section{Conclusion}\label{S5}

In this work, we have theoretically investigated the dynamic response of
a self-gravitating Bose--Einstein condensate (BEC) dark matter (DM) halo
with axion-like self-interaction~\cite{Naz2025,GN23} to subgalactic
baryonic perturbations.
By confining our hydrodynamic and spectral calculations to the central
galactic region of the dwarf galaxy NGC 2366, which overlaps with the
stable DM core, we preserve spherical symmetry while extracting
analytically traceable physical trends applicable to general galactic
configurations.

Specifically, our linearized hydrodynamic analysis delineates the structural
response of the DM halo core under realistic baryonic feedback. By modeling
the stellar and gaseous components via tailored multi-cluster Plummer shell
distributions parametrized by empirical data for NGC~2366, we tracked the
expansion, collapse, and migration of baryonic matter. Notably, the stellar
migration scenario incorporates an extension of the classical Chandrasekhar
dynamical friction formula adapted to our concentric two-shell geometry,
thereby capturing the supersonic drag of the cluster embedded within the
gaseous environment.

The resulting DM responses reveal that despite the presence of an anomalous
compressibility zone ($c_s^2<0$), restricting the geometry to the central
galactic region guarantees core stabilization through spatial isolation from
global halo modes. The simulations demonstrate that while central density
variations remain subtle, the perturbation peaks expand to kiloparsec scales
and adhere to a strict inverse relationship, where a higher density accumulation
suppresses the local flow velocity. Most remarkably, these baryon-induced DM winds
are shown to be highly persistent, remaining unattenuated long after the baryonic
drive has relaxed, which complicates baseline assumptions for direct DM detection
experiments.

Furthermore, we have evaluated the fate of cumulative, cascaded
perturbations through a quantum spectral approach. By replacing temporary
hard-wall constraints with realistic Robin boundary conditions, we modeled
the smooth matching between the dense galactic core and the extended outer
halo tail. The resulting non-equidistant discrete spectrum unveils a unique
quantum relaxation mechanism operating in the absence of physical viscosity.
The lowest long-wavelength mode possesses a purely imaginary frequency,
characterizing a metastable false vacuum state localized within the sub-Jeans
core region. Quantum tunneling of fluctuations across the self-generated
barrier induces a collective, non-unitary relaxation with a finite decay
timescale of $6.8 \times 10^7$ years, which is well within the cosmological
age of dwarf galaxies. In higher perturbation orders, the breakdown of
continuous energy-momentum conservation restricts standard Beliaev damping,
giving way to a nonlinear three-mode parametric wave mixing that cascades
energy toward stable long-wavelength excitations.

Ultimately, these findings complement and extend the broader framework of
self-interacting DM evolution~\cite{MSch15,Sch23,RG05,Mash06,GM06,Gov10,Gov12,KL22}.
While individual baryonic events cause minor
immediate density feedback, their cumulative quantum relaxation plays a
critical role in regulating the macroscopic structural stability of BEC DM
halos. Future extensions of this model will incorporate non-spherical
configurations and evaluate the cross-mode nonlinear stabilization of
higher angular multipoles to further detail the lifetime of DM cores
under external cosmological influences.

%%%%%%%%%%%%%%%%%%%%%%% Acknowledgments %%%%%%%%%%%%%%%%%%%%%%%
\section*{Acknowledgments}

This work is supported by the Ukrainian-Swiss project No. IZURZ2-224866
``Compact star-forming galaxies and their impact on cosmic reionization
and cosmology'' (\url{https://data.snf.ch/grants/grant/224866}).

%% %% %% %% %% %% %% %% %% %% %% %% %% %% %% %% %% %% %% %% %% %% %% %% %% %% %% %%
\appendix

\section{Surface density}\label{ApA}

Here we focus on obtaining the surface density of a unit mass
by starting from the integral:
\begin{equation}
\Sigma_0(R,\breve{r},d)=2\int_{0}^\infty\eta_{\rm b}\left(\sqrt{z^2+R^2},\breve{r},d\right)
\,\rmd z.
\end{equation}

Defining the parameters $a=\breve{r}/R$ and $b=d/R$, we first
compute the auxiliary elliptic integral of the first kind
(see formulas 3.145 in \cite{Tbls}):
\begin{eqnarray}
I_1(a,b)&\equiv&\int_0^1
\left[\frac{1}{\sqrt{(1-a s)^2+b^2 s^2}}-\frac{1}{\sqrt{(1+a s)^2+b^2 s^2}}\right]
\frac{\rmd s}{\sqrt{1-s^2}}
\nonumber\\
&=&\frac{2}{\sqrt{y}}\left[K(k)-F(x;k)\right],
\end{eqnarray}
which is determined by
\begin{eqnarray}
x&=&\frac{2\sqrt{y}}{\sqrt{(1+a)^2+b^2}+\sqrt{(1-a)^2+b^2}},
\nonumber\\
k&=&\frac{\sqrt{a^2+b^2-1+y}}{\sqrt{2y}},
\label{yxk}\\
y&=&\sqrt{[(1+a)^2+b^2]\,[(1-a)^2+b^2]},
\nonumber
\end{eqnarray}
where $x$ is the sine of the amplitude, $k$ is the modulus.

The desired function is represented through the derivative of $I_1(a,b)$:
\begin{eqnarray}
I_2(a,b)&\equiv&-\frac{\partial I_1(a,b)}{b\,\partial b}
\nonumber\\
&=&\frac{1}{\sqrt{y}}\left(\frac{y_b}{y}+2\frac{k_b}{k}\right)
\left[K(k)-F(x;k)\right]
+\frac{2}{\sqrt{y}}\frac{x_b}{\sqrt{1-x^2}\sqrt{1-k^2x^2}}
\nonumber\\
&&
-\frac{2k_b}{k (1-k^2)\sqrt{y}}\left[E(k)-E(x; k)
+k^2\frac{x\sqrt{1-x^2}}{\sqrt{1-k^2x^2}}\right],
\end{eqnarray}
where $E(k)$ and $E(x; k)$ are the complete and incomplete elliptic
integrals of the second kind; the subscript $b$ for each parameter
$w$ from Eq.~(\ref{yxk}) denotes the differentiation $w_b=b^{-1}\partial_bw$.

Calculating, we find that
\begin{equation}
x_b=\frac{x}{y^2}(1+a^2+b^2-y),\quad
k_b=\frac{1+b^2-a^2}{y^3k},\quad
y_b=2\frac{1+a^2+b^2}{y}.
\end{equation}

Combining, we obtain the surface density profile:
\begin{equation}
\Sigma_0(R,\breve{r},d)=\frac{d^2}{\breve{r} R^3}
\,I_2\left(\frac{\breve{r}}{R},\frac{d}{R}\right).
\end{equation}
This distribution is normalized as
\begin{equation}
\int_0^{2\pi}\rmd\varphi\int_0^\infty \Sigma_0(R,\breve{r},d)\, R\rmd R=4\pi.
\end{equation}

Thus, the dimensional spatial and surface distributions for the spherical layer (cluster)
with the total baryon mass $M$ are
\begin{eqnarray}
&&\rho_M(r,\breve{r},d)=\frac{M}{4\pi}\eta_{\rm b}(r,\breve{r},d),
\label{spt}\\
&&\Sigma_M(R,\breve{r},d)=\frac{M}{4\pi}\Sigma_0(R,\breve{r},d).
\label{srf}
\end{eqnarray}

%% %% %% %% %% %% %% %% %% %% %% %% %% %% %% %% %% %% %% %% %% %% %% %% %% %% %% %%

\end{document}